\documentclass[3p,10pt,numbers]{elsarticle}
\usepackage{graphicx}
\usepackage{amsmath,amssymb}
\usepackage{upgreek,fixmath}
\usepackage{color}
\usepackage[hidelinks]{hyperref}
\usepackage{txfonts}

\newcommand{\rcoord}{\ensuremath{r}}
\newcommand{\fourier}{\ensuremath{\mathcal{F}_\bb{k}}}
\newcommand{\spaceOperator}{\ensuremath{\hat{\mathcal{X}}}}
\newcommand{\rarg}{(\rcoord_0)}
\newcommand{\tor}{\zeta}
\usepackage{pifont}

\usepackage{macros}

\newcommand{\dblbrck}[1]{[#1]}

\journal{Journal of Computational Physics}

\title{A novel approach to radially global gyrokinetic simulation using the flux-tube code  \texttt{stella} }

  \author[add1]{D. A. St-Onge\corref{cor1}}
  \ead{denis.st-onge@physics.ox.ac.uk}
  \author[add1,add2]{M. Barnes}
  \author[add4]{F.I. Parra}

  \cortext[cor1]{Please address correspondence to denis.st-onge@physics.ox.ac.uk}
  \address[add1]{Rudolf Peierls Centre for Theoretical Physics, University of Oxford, Oxford OX1 3PU, United Kingdom}
  \address[add2]{University College, Oxford OX1 4BH, United Kingdom}
    \address[add4]{Princeton Plasma Physics Laboratory, Princeton University, Princeton, New Jersey 08544}

\date{\today}

\begin{document}
\begin{frontmatter}


\begin{abstract}
    A novel approach to global gyrokinetic simulation is implemented in the flux-tube code~\texttt{stella}. This is done by using a subsidiary expansion of the gyrokinetic equation in the perpendicular scale length of the turbulence, originally derived by
    Parra and Barnes \dblbrck{Plasma Phys.~Controlled Fusion, \textbf{57} 054003, 2015}, which allows the use of Fourier basis functions while enabling the effect of radial profile variation to be included in a perturbative way. Radial variation of the magnetic geometry is included by utilizing a global extension of the Grad-Shafranov equation and the Miller equilibrium equations which is obtained through Taylor expansion.
    Radial boundary conditions that employ multiple flux-tube simulations are also developed, serving as a more physically motivated replacement to the conventional Dirichlet radial boundary conditions that are used in global simulation. It is shown that these new boundary conditions eliminate much of the numerical artefacts generated near the radial boundary when expressing a non-periodic function using a spectral basis. We then benchmark the new approach both linearly and non-linearly using a number of standard test cases.

\end{abstract}

\end{frontmatter}

\section{Introduction}

Numerical simulation has been an important tool for plasma physics and fusion research since the inception of particle-in-cell methods in the 1950s~\citep{BirdsallLangdon}. Subsequent developments, such as the formulation of the gyrokinetic system of equations~\citep{Catto_lin,FriemanChen,Dubin} and flux-tube simulations that employ magnetic-field-following coordinates~\citep{Beer95}, have enabled computationally affordable numerical studies of  otherwise enormously complex physical systems. These studies have been instrumental in various discoveries, such as the importance of zonal flows in reducing turbulent transport in axisymmetric systems~\citep{Dimits} and novel plasma modes~\citep{Hallatschek_tails,Parisi_modes}. Gyrokinetic simulations have also successfully matched turbulent heat fluxes from experimental data~\citep{Candy_2003}, have shown the importance of interactions between the ion and electron scales~\citep{Howard_2016}, and have been used to predict new phenomena that have been borne out in experiment~\citep{DiSiena_prediction}. For a review on the history and state of the art of plasma simulation for fusion science, see~\citet{White_2017} and references therein.

Gyrokinetic simulation of tokamak and stellarator plasmas can be divided into two distinct but related paradigms, that of `global' and `local' simulation. In the global paradigm, which was the original approach used in early gyrokinetic studies~\citep{Parker93}, the entire physical device (or a finite portion thereof) is modelled, which includes the spatial variation of the magnetic geometry and pressure profiles which make up the equilibrium state. This approach has the advantage of being able to capture effects arising from radial profile variation, such as turbulence spreading~\citep{Garbet_1994}.
 However, due to radial variation in the equilibrium temperature and magnetic geometry, the gyro-averaging operator becomes spatially dependent, and thus difficult to calculate; early studies were forced to adopt simplifications that made calculation of the gyroaverage tractable. More sophisticated, non-spectral gyroaveraging schemes  have also been developed~\citep{Jolliet_2007,Steiner_2015}, but these schemes either suffer from low accuracy or high computational cost~\citep{guadagni_cerfon_2017}. Additionally, global simulations typically use Dirichlet boundary conditions which nullify perturbations at the radial boundaries. With sufficiently large `buffer' regions at the radial boundaries, such boundary conditions are hoped to be `benign', though it is unclear what effects they have on physical observables without detailed studies for each simulation~\citep{Candy_globloc}. The `local' paradigm stems from the development of the flux-tube domain~\citep{Beer95}, which considers an infinitesimally thin box elongated along a magnetic field line. As the flux-tube samples the equilibrium profiles at a specific radial location, the gyrokinetic equation becomes homogeneous in the direction perpendicular to the magnetic field, and so statistical periodicity can be assumed, allowing the use of Fourier spectral methods and thereby permitting an analytical treatment of the gyroaveraging operator. This development led to a number of fixed-grid gyrokinetic codes~(e.g.,~\citep{gyro,gs2,gene}) that can probe the local characteristics of turbulence with high fidelity at only moderate computational costs. This, however, came at the expense of neglecting any influence from the effects of radial profile variation.

Formally, local gyrokinetics is the $(k_\perp L)^{-1} \rightarrow 0$ limit of global gyrokinetics, and numerical studies have indeed shown this to be the case~\citep{Candy_limit}. Here, $k_\perp$ is the inverse scale length of the fluctuations across the magnetic field and $L$ is a characteristic large-scale size (such as the device minor radius $a$). Recently, it has been shown by~\citet{Parra_globallocal} that the local and global approaches to gyrokinetic simulation can be bridged using an appropriate expansion of the gyrokinetic equation, and that an approach that sits between the local and global paradigms can be formulated. Indeed, radial profile variation of the pressure has been recently implemented in the flux-tube code CGYRO using periodic triangle waves~\citep{Candy_globshear,Candy_globloc}. In this manuscript, we formulate a new approach to hybrid global-local gyrokinetics that is related to the one laid out in~\citep{Parra_globallocal}, which includes novel boundary conditions, equilibrium flow shear algorithms, and sources and sinks. We then implement our approach using an extended version of the gyrokinetic flux-tube code~\texttt{stella}~\citep{Barnes_stella}, benchmark it against a number of standard linear and nonlinear test cases, and discuss some limitations with regards to the novel radial boundary conditions.

This paper is organized as follows: in \S \ref{sec:theory} we describe the gyrokinetic equation and lay out the theoretical framework that underlies our approach to global gyrokinetics. The numerical implementation of our approach in the flux-tube code \texttt{stella} is explained in~\S \ref{sec:numerical_imp}, with additional details on the profile variation of magnetic geometry, the moments of the distribution function and the fluxes given in~\ref{app:mag_geo} and \ref{app:fluxes}. In \S \ref{sec:tests} we perform benchmarks of our global approach and compare it to \texttt{stella} in local operation. We offer concluding remarks in~\S \ref{sec:conclusion}.


\section{Theoretical Model}\label{sec:theory}

\subsection{Gyrokinetic equation}\label{sec:GKE_delta}

The \texttt{stella} code numerically solves the first-order $\delta f$ gyrokinetic equation, which is derived by taking the Vlasov equation and then imposing the ordering 
\begin{equation}\label{eqn:orderings}
 \epsilon \sim \rho_\ast  \sim \frac{\delta f_s}{f_s } \sim \frac{\omega}{\Omega_\mr{i}} \sim \frac{k_\parallel}{k_\perp} \sim \frac{Z_s e\varphi}{T_s}\ll 1,
\end{equation}
where $\epsilon$ is the ordering parameter, $\rho_\ast \doteq \rho_\mr{ref} / L$, $\rho_\mr{ref}$ is a reference thermal gyroradius where $\rho_\mr{ref}\sim \rho_\mr{i}$ and $\rho_\mr{i} = v_\perp / \Omega_\mr{i}$ is the ion gyroradius, $v_\perp$ is the velocity perpendicular to the magnetic field,
$f_s$ and $\delta f_s$ are respectively the total and perturbed distribution function for species $s$, $\omega$ is a characteristic frequency of the perturbations, $\Omega_\mr{i}$ is the ion gyrofrequency,  
$k_\parallel$ is the characteristic inverse scale length of the perturbations along the magnetic field, $Z_s$ is the charge number, $e$ is the (positive) unit charge, $\varphi$ is the electrostatic potential and $T_s$ is the temperature for species $s$.
This ordering allows one to average over the fast gyromotion of charged particles in a strong magnetic field, which removes the dependence of the particle gyrophase. Thus, the dimensionality of phase space is reduced to three spatial dimensions and two velocity dimensions. Additionally, the fast gyrofrequency timescale is removed from the system. Finally, it is assumed that $f_s = F_s + \delta f_s$, where $F_s$
is a stationary background Maxwellian distribution given by
\begin{equation}
F_s = n_s\left(\frac{m_s}{2\upi T_s}\right)^{3/2}\exp\left( - \frac{m_s v_\parallel^2}{2T_s} - \frac{\mu_s B}{T_s}\right).
\end{equation}
Here, $s$ is the species index; $m_s$ and $n_s$ are the particle mass and density for species $s$, respectively; $v_\parallel = \eb \bcdot \bb{v}$;  $\eb$ is the unit vector along the magnetic field; $\bb{v}$ is the peculiar velocity in the frame rotating with the plasma at the center of the radial domain, $\mu_s = m_s v_\perp^2 / 2B $ is the magnetic moment; and $B$ is the magnitude of the magnetic field. Neglecting electromagnetic effects and collisions, and considering only toroidal flow~\citep{Catto_flow}, the gyrokinetic equation is given by
\begin{align}\label{eqn:GKE}
\frac{\partial g_s}{\partial t} & +R\Omega_\tor \bb{\hat{\tor}}\bcdot \grad g_s + v_\parallel  \left(\underline{\eb \bcdot \grad } g_s  + \underline{\frac{Z_se}{T_s} F_s \eb \bcdot \grad }\langle \varphi \rangle_\bb{R} \right)- \frac{\mu_s}{m_s} \underline{\eb \bcdot \grad B} \frac{\partial g_s}{\partial v_\parallel} \nonumber \\ &+ \underline{\bb{v}_{\mr{M}s}}\bcdot \left(\grad_\perp g_s +\underline{\frac{Z_se}{T_s}F_s}\grad_\perp \langle \varphi\rangle_\bb{R}\right) + \underline{\langle\bb{v}_\bb{E} \rangle_\bb{R} \bcdot \grad_\perp} g_s + \underline{\langle \bb{v}_\bb{E} \rangle_\bb{R}\bcdot \grad\big|_E F_s} +  \underline{\frac{m_s }{T_s}F_s\frac{Iv_\parallel}{B}\langle\bb{v}_\bb{E} \rangle_\bb{R} \bcdot \grad_\perp} \Omega_\tor = 0, 
\end{align}
where $g_s(\bb{R},v_\parallel, \mu_s, t) \doteq \ba{\delta f}_\bb{R}$ is the perturbed gyrocenter distribution function for species $s$, underlines indicate coefficients that have spatial dependence,  $\langle \cdots \rangle_\bb{R}$ denotes gyroaveraging at constant gyrocenter position $\bb{R} = \bb{r} - \bb{\rho}$, $\bb{\rho} = \eb \btimes \bb{v}/\Omega_s$ is the gyroradius vector, $\Omega_s$ is the gyrofrequency for species $s$, $R$ is the device major radius, $\bb{\hat{\tor}}$ is the toroidal unit vector, $I = RB_\mr{T}$ where $B_\mr{T}$ is the toroidal magnetic field strength, $\Omega_\tor$ is the toroidal rotation which varies across flux surfaces,  
$\grad|_\mr{E}$ is the gradient taken at constant kinetic energy $E=m_s v^2 / 2$, the magnetic and $\bb{E}\btimes \bb{B}$ drifts are given by
\begin{subequations}
\begin{align}
\bb{v}_{\mr{M}s} &= \frac{\eb}{\Omega_s}\btimes \left(\frac{\mu_s}{m_s} \grad B + v_\parallel^2 \bb{\kappa}\right), \\
\bb{v}_\bb{E} &= \frac{c}{B}\eb \btimes \grad_\perp \varphi,
\end{align}
\end{subequations}
respectively, $c$ is the speed of light,  $\bb{\kappa} = \eb \bcdot \grad \eb$ is the magnetic field curvature and $\bb{E}$ and $\bb{B}$ are the electric and magnetic fields, respectively. The gyrokinetic equation~\eqref{eqn:GKE} is boosted to the frame moving along the mean toroidal flow $R \Omega_\tor$.  This flow is assumed to be subsonic, i.e., $R\Omega_\tor \sim \epsilon v_\mr{thi}$, and is allowed to have a strong flow gradient, i.e. $R^2 \Omega'_\tor \sim v_\mr{thi}$, where prime denotes differentiation with respect to $\rcoord$.
This ordering implies that the Coriolis and centrifrugal drifts, given by
\begin{subequations}
\begin{align}
    \bb{v}_{\mr{Coriolis},s} &=  \frac{2 v_\parallel \Omega_\tor}{\Omega_{s}} \eb \btimes [(\grad_\bb{R} R \btimes \bb{\hat{\zeta}}) \btimes \eb]\\
    \bb{v}_{\mr{centrifugal},s} &= - \frac{R\Omega^2_\tor}{\Omega_s}\eb\btimes \grad_\bb{R}R
\end{align}
\end{subequations}
are respectively ordered $\rho_\ast$  and $\rho_\ast^2$ compared to the magnetic drifts, and so are not included in equation~\eqref{eqn:GKE}, while the terms dealing with perpendicular and parallel flow shear are retained. 
In order to close equation~\eqref{eqn:GKE}, $\varphi$ must be determined. This is done using the quasineutrality equation
\begin{equation}\label{eqn:quasi}
    \sum_s Z_s \delta n_s \doteq \sum_s Z_s \int \od^3 v\left(\langle g_s\rangle_\bb{r} + \frac{Z_s e}{T_s} F_s \left(\langle\langle \varphi\rangle_\bb{R}\rangle_\bb{r} - \varphi\right)\right) =0,
\end{equation}
where $\delta n_s$ is the perturbed gyrocenter density for species $s$, $\langle \cdots \rangle_\bb{r}$ is the gyroaverage taken at constant particle position $\bb{r}$ and the Debye length is taken to be much smaller than the electron gyroradius.

The first order $\delta f$ gyrokinetic equation~\eqref{eqn:GKE} forms the basis of both conventional global gyrokinetics and local flux-tube gyrokinetics, the latter being the $(k_\perp L)^{-1} \rightarrow 0$ limit of the former~\citep{Parra_globallocal}. To be more precise, for global gyrokinetics the geometrical coefficients and background pressure profiles \dblbrck{the underlined terms in~\eqref{eqn:GKE}} are allowed to vary in all spatial directions. In addition, the gyroaveraging operator may also have spatial dependence.  As $(k_\perp L)^{-1} \rightarrow 0$, these terms no longer vary across turbulent eddies and can thus be treated as constants in the directions perpendicular to the magnetic field; the only variation of these quantities that persists is along the magnetic field. Local gyrokinetics typically assumes periodicity in a statistical sense in the perpendicular directions, and thus local flux-tube gyrokinetics is routinely performed using a pseudo-spectral approach on a Fourier basis and field-line-following Clebsch coordinates $\alpha$ and $\psi$, where $\psi$ is a flux label, $\alpha$ labels a fieldline, and $\bb{B} = \grad \alpha \btimes \grad \psi$. The coordinate along the direction of the magnetic field, denoted by $z$, is typically chosen in a way to simplify the underlying equations; some examples are the arc length along the magnetic field line, the toroidal angle $\zeta$ or the poloidal angle $\theta$.

\subsection{Subsidiary expansion of the gyrokinetic equation \label{sec:subsidiary}}

In practice, while local simulations employ the $(k_\perp L)^{-1} \rightarrow 0$ limit, the extent of their radial domains are typically the size of hundreds of gyroradii and thus cannot be considered `small' when compared to typical device parameters; it is this largeness of the radial box size that we exploit in order to incorporate profile variation in local flux-tube gyrokinetics. Defining the parameter $\Delta \doteq \ell_x/L$, where $\ell_x$ is the radial extent of a simulation, we now impose a subsidiary expansion on the gyrokinetic equation~\eqref{eqn:GKE} using the ordering  
\begin{equation}\label{eqn:subsidiary}
 \epsilon \sim \frac{1}{k_\perp L}  \ll  \Delta \ll 1.
\end{equation}
 This ordering allows us to include the effects of profile variation without needing to go to higher order in $\rho_\ast$.\footnote{\label{footnote:rhostar}It is worthwhile to point out that most global gyrokinetic codes do not retain all higher order terms in $\rho_\ast$, e.g., diamagnetic flows~\citep{Lee_NF, Lee_POP, Lee_PPCF, Barnes_diamagnetic} and the parallel nonlinearity~\citep{Kniep_2004, Lin_parallel,McDevitt_2009, Candy_parallel,Barnes_parallel}} 

In this manuscript, we focus on the radial profile variation found in axisymmetric geometry which is relevant to tokamaks. We defer the treatment of the `full flux surface' version of \texttt{stella}, which includes azimuthal variation found in three-dimensional systems relevant to stellarators, to a later  publication. For tokamak geometry and equilibria, axisymmetry implies that, while we only consider  profile variation along the magnetic field and across flux surfaces, radially global \texttt{stella} is also toroidally global, without needing any further consideration of the binormal direction $\eb \btimes \grad \psi$ (the direction perpendicular to the magnetic field lying within a flux surface). In order to make \texttt{stella} fully global, one must also include higher order effects in the parallel physics, such as the parallel nonlinearity (see references in footnote~\ref{footnote:rhostar}) and higher order parallel derivatives in the particle drifts~\citep{Sung_2013}; these extra terms are not considered in this manuscript.

To perform the subsidiary expansion given in~\eqref{eqn:subsidiary}, we first normalize the gyrokinetic equation to render it dimensionless. In general there are two ways to do this: have the normalizing quantities vary across flux surfaces or have them fixed to a reference surface at some radial position $r_0$. While the former is the most economical in terms of the velocity space resolution required, it introduces additional derivatives to the gyrokinetic equation (including with respect to $\mu_s$) as well as complicating the multiple flux-tube boundary conditions given in \S \ref{sec:radBC}. The \texttt{stella} code is  efficiently parallelized over velocity space, allowing for large velocity-space grids, and so we opt for normalization at a fixed reference position. 
Denoting reference values using a subscript `ref' and normalized quantities with a tilde, the normalized quantities are  $\tilde{\bb{v}} = \bb{v} / v_{\mr{th}s}(\rcoord_0)$, $\tilde{v}_{\parallel} = v_\parallel/v_{\mr{th}s}(\rcoord_0)$, $\tilde{\mu}_s = \mu_s B_\mr{ref}/m_s v_{\mr{th}s}^2(\rcoord_0)$, $\tilde{B}= B/B_\mr{ref}$, $\tilde{m}_s = m_s/m_\mr{ref}$, $\tilde{n}_s = n_s/n_\mr{ref}$, $\tilde{T}_s = T_s/T_\mr{ref}$, $\tilde{I} = RB_\mr{T}/aB_\mr{ref}$, and $\tilde{\Omega}_\zeta = a \Omega_\zeta/v_\mr{th,ref}$. Here, $v_{\mr{th}s} = \sqrt{2 T_s(\rcoord) / m_s}$.
We also normalize time to $a/ v_{\mathrm{th,ref}}$, parallel lengths to $a$ and perpendicular lengths to $\rho_\mr{ref}$, where $a$ is the minor radius and $v_{\mathrm{th,ref}}=\sqrt{2T_\mr{ref}/m_\mr{ref}}$,  $\rho_\mr{ref}= v_\mr{th,ref}/\Omega_\mr{ref}$ and $\Omega_\mr{ref} = eB_\mr{ref}/m_\mr{ref}c$. 
Finally, we introduce the normalized  distribution function and electrostatic potential $\tilde{g}_s = (a/\rho_\mr{ref})[g_s/F_s(\rcoord_0)]\exp[-v^2(\rcoord_0)/v_{\mr{th}s}^2(\rcoord_0)]$ and $\tilde{\varphi} =  (e\varphi/T_\mr{ref})(a/\rho_\mr{ref})$, where $v^2(\rcoord) = v_\parallel^2 + 2B(\rcoord) \mu_s/m_s$. The normalized gyrokinetic equation is then
 \begin{align}\label{eqn:gk_norm}
\frac{\partial \tilde{g}_s}{\partial \tilde{t}} &+ \frac{R}{a}\tilde{\Omega}_\zeta \bb{\hat{\tor}}\bcdot \tilde{\grad} \tilde{g}_s + \frac{v_{\mr{th}s}(\rcoord_0)}{v_{\mr{th,ref}}}\tilde{v}_\parallel \eb \bcdot \tilde{\grad} \tilde{z} \left(\frac{\partial \tilde{g}_s}{\partial \tilde{z}}  +  \frac{\partial \langle\tilde{\varphi} \rangle_\bb{R} }{\partial \tilde{z}} \frac{Z_s}{\tilde{T}_s(\rcoord)}\frac{F_s(\rcoord)}{F_s(\rcoord_0)} \rme^{-{v^2(\rcoord_0)}/v_{\mr{th}s}^2(\rcoord_0)} \right) - \frac{v_{\mr{th}s}(\rcoord_0)}{v_\mr{th,ref}}\tilde{\mu}_s \eb \bcdot \tilde{\grad} \tilde{B} \frac{\partial \tilde{g}_s}{\partial \tilde{v}_\parallel}  \nonumber
 \\ & + \tilde{\bb{v}}_{\mr{M}s}\bcdot \left(\tilde{\grad}_\perp \tilde{g}_s +\tilde{\grad}_\perp \langle\tilde{\varphi} \rangle_\bb{R}\frac{Z_s}{\tilde{T}_s(\rcoord)}\frac{F_s(\rcoord)}{F_s(\rcoord_0)}\rme^{-{v^2(\rcoord_0)}/v_{\mr{th}s}^2(\rcoord_0)} \right)+  \langle\tilde{\bb{v}}_\bb{E} \rangle_\bb{R} \bcdot \tilde{\grad}_\perp \tilde{g}_s    +\frac{\rme^{-{v^2(\rcoord_0)}/v_{\mr{th}s}^2(\rcoord_0)} }{F_s(\rcoord_0)}  \langle\tilde{\bb{v}}_\bb{E} \rangle_\bb{R} \bcdot  a\grad\big|_E F_s\nonumber \\
 &  +  {\frac{\tilde{m}_s }{\tilde{T}_s(\rcoord)} \frac{v_{\mr{th}s}(\rcoord_0)}{v_\mr{th,ref}}\frac{F_s(\rcoord)}{F_s(\rcoord_0)}\rme^{-{v^2(\rcoord_0)}/v_{\mr{th}s}^2(\rcoord_0)}\frac{\tilde{I}\tilde{v}_\parallel}{\tilde{B}}\langle\tilde{\bb{v}}_\bb{E} \rangle_\bb{R} \bcdot \tilde{\grad}_\perp} \tilde{\Omega}_\tor = 0, 
\end{align}
where 
\begin{subequations}
\begin{align}
    \tilde{\bb{v}}_{\mr{M}s} &=  \frac{T_s(\rcoord_0)}{Z_sT_\mr{r}}\frac{1}{\tilde{B}(\rcoord)} \eb\btimes \left(\tilde{\mu}_s\tilde{\grad} \tilde{B} + \tilde{v}_\parallel^2 \tilde{\bb{\kappa}}\right), \\
      \tilde{\bb{v}}_{\bb{E}} &=  \frac{1}{2} \frac{1}{\tilde{B}(\rcoord)}\eb \btimes \tilde{\grad}_\perp \tilde{\varphi}.
\end{align}
\end{subequations}
The quasineutrality equation is also normalized,
\begin{equation}\label{eqn:quasi_norm}
   \sum_s Z_s \tilde{n}_s(\rcoord_0) \int \od^3 \tilde{v}\frac{1}{\upi^{3/2}}\left(\langle \tilde{g}_s\rangle_\bb{r} + \frac{Z_s }{\tilde{T}_s(\rcoord)} \frac{F_s(\rcoord)}{F_s(\rcoord_0)} \rme^{-{v^2(\rcoord_0)}/v_{\mr{th}s}^2(\rcoord_0)} \left(\langle\langle \tilde{\varphi}\rangle_\bb{R}\rangle_\bb{r} - \tilde{\varphi}\right)\right) =0.
\end{equation}

\subsubsection{Radial variation of pressure and magnetic geometry \label{sec:var_outline}}

The expansion in $\Delta$ outlined in \S \ref{sec:GKE_delta} allows us to incorporate variation of the equilibrium pressure profiles into global \texttt{stella} by considering their Taylor expansions around a central radial location $r_0$:
\begin{subequations}
\begin{align}
n_s(r) = n_s(r_0) + \left. \frac{\od n_s}{\od r}\right|_{r=r_0}(r-r_0) + 
\frac{1}{2}\left. \frac{\od^2 n_s}{\od r^2}\right|_{r=r_0}(r-r_0)^2 + \ldots , \\ 
T_s(r) = n_s(r_0) + \left. \frac{\od T_s}{\od r}\right|_{r=r_0}(r-r_0) + 
\frac{1}{2}\left. \frac{\od^2 T_s}{\od r^2}\right|_{r=r_0}(r-r_0)^2 + \ldots .
\end{align}
\end{subequations}
The inclusion of the new terms $n_s''(r)$ and $T''_s(r)$, where primes denote differentiation with respect to $\rcoord$, modifies the  source of free energy resulting from the gradients of the background distribution function $\left.\grad\right|_EF_s$.  Additionally, the expansion in $\Delta$ also results in $n'$ and $T'$ modifying  both the magnetic drift and parallel streaming terms through the adiabatic contribution $(Z_s/T_s)F_s \varphi$  of the perturbed distribution function. These first derivatives also introduce profile variation of the quasineutrality equation~\eqref{eqn:quasi}.

While the density profile is specified by $n_s(\rcoord_0)$, 
$n'_s(\rcoord_0)$, $n''_s(\rcoord_0)$ (similarly for the temperature profiles),  more relevant in determining growth rates and turbulent amplitudes are the gradient scale lengths $L_{n_s}^{-1} = - n'_s/n_s$ and $L_{T_s}^{-1} = - T'_s/T_s$. While global \texttt{stella} utilizes the second derivatives of the density and temperature, these are not required for  profile variation of the gradient scale lengths:
\begin{equation}
\left(L_n^{-1}\right)' =\frac{n'^2_s}{n^2_s} -\frac{n''_s}{n_s},
\end{equation}
 likewise for temperature, and so profile variation of the growth rates and turbulent amplitudes is expected, even with $n_s'' = T_s'' = 0$.


Along with radial variation of the background pressure profiles, global \texttt{stella} also includes variation of the magnetic geometry through the use of an extended Miller equilibrium model~\citep{miller}. The original Miller equilibrium parameterises a single axisymmetric flux surface using the model equations
\begin{subequations}\label{eqn:miller}
\begin{align}
R(r,\theta) &= R_0(r) + r \cos (\theta + \sin \theta \arcsin \delta (r)), \\
Z(r,\theta) &= \kappa(r) r \sin (\theta), 
\end{align}
\end{subequations}
where $R$ and $Z$ are the radial and vertical position in cylindrical coordinates, $r$ is the flux surface label, $\theta$ is a poloidal coordinate, $R_0$ may include the radial displacement of a given flux surface (Shafranov shift), and $\delta(r)$ and $\kappa(r)$ set the triangularity and elongation of the flux surface, respectively. In order to calculate the geometrical coefficients of~\eqref{eqn:GKE}, one must first locally solve the Grad-Shafranov equation. This requires specification of the derivatives $R'_0(r)$, $\delta'(r)$, and $\kappa'(r)$, as well as specification of the safety factor $q$ \dblbrck{defined in \eqref{eqn:app_q}}, the normalized plasma pressure $\beta= 4 \upi p/B_\mr{ref}^2$, and their derivatives. Here, $p$ is the species-summed plasma pressure.

To extend the Miller equilibrium model to a global profile, one must ensure that the Grad-Shafranov equation is consistently satisfied over the entire profile; it is not sufficient to evaluate~\eqref{eqn:miller} at different radial locations $r$ and then to use these coefficients to individually solve the Grad-Shafranov equation at these locations. Global \texttt{stella} instead solves the Grad-Shafranov equation localized at some specified central radius $r=r_0$, and then also its first radial derivative centered at the same location. This ensures that the geometrical coefficients evolve across the radial domain in a way consistent with MHD equilibria and with the subsidiary expansion in $\Delta$ outlined in \S \ref{sec:GKE_delta}. This approach requires the specification of the second radial derivatives $q''(r)$, $\psi''(r)$ and $\beta''(r)$. Like with the density and the temperature, a non-zero value $q''(r)$ is not required for profile variation in the shearing parameter $\hat{s} \doteq (r/q) q'(r)$:
\begin{equation}
   \hat{s}' = \frac{r q''}{q} + \frac{q'}{q} - \frac{r q'^2}{q^2}.
\end{equation}
 The coefficients of~\eqref{eqn:gk_norm} that depend on $\rcoord$ are now Taylor expanded up to first order, and the resulting equations are given in \S \ref{sec:sim_equations}. Detailed calculations of the geometrical coefficients and their derivatives are given in \ref{app:mag_geo}.

\subsection{Coordinates and the parallel boundary condition}\label{sec:coords}

The coordinates used in \texttt{stella} for axisymmetric systems are $(x,y,\theta,v_\parallel,\mu_s)$, where $\theta$ is a poloidal coordinate which signifies the position along a magnetic field line (as used in the Miller equilibrium) and is zero at the outboard midplane.\footnote{The reader can refer to~\citet{Barnes_stella} for the use of \texttt{stella} in three-dimensional systems.} The $(x,y)$ coordinates are in the plane perpendicular to $\bb{B}$, with the binormal coordinate $y$ being defined in the same way for both the local and global versions of \texttt{stella}:
\begin{equation}\label{eqn:ycoord}
y = \frac{\psi'\rarg}{B_\mr{ref}}(\alpha-\alpha_0),
\end{equation}
where $\psi$ is the poloidal flux function, $B_\mr{ref}$ is a reference magnetic field, $\alpha = \tor - q \vartheta$, $\tor$ is the toroidal angle and $\vartheta$ is the straight-field-line poloidal angle, given by~\eqref{eqn:vartheta_app}.

Currently, \texttt{stella} offers two options for the choice of the radial coordinate $x$, the first being the poloidal flux function $\psi$,
\begin{equation}\label{eqn:xcoord_loc}
x = \frac{q_0}{r_0 B_\mr{ref}}(\psi-\psi_0),
\end{equation}
while the second is the safety factor $q$,
\begin{equation}\label{eqn:xcoord_glob}
x = \frac{\psi'\rarg}{B_\mr{ref}} (q-q_0).
\end{equation}
For local \texttt{stella}, $\psi$ is used as the radial coordinate by default. Conversely, global \texttt{stella} is made to use $q$ as the radial coordinate in order to simplify the implementation of the parallel boundary condition typically used in flux-tube simulations~\cite{Beer95}. This boundary condition, which is for the $\theta$ dimension,\footnote{While $\theta$ is used for the parallel coordinate, $\vartheta$ is the variable that appears in the definition of $\alpha$. The angles $\theta$ and $\vartheta$ can be made to match at the inboard midplane ($\vartheta(\theta = \uppi) = \upi$ and $\vartheta(\theta = -\upi) =-\upi$ ), the location at which the boundary condition is applied in simulation. Therefore, either variable can be used in the parallel boundary condition.} asserts $2\upi N$ periodicity of a quantity $A$ at fixed toroidal angle $\tor$, rather than for fixed $\alpha$,
\begin{equation}\label{eqn:parBC}
    A(\psi, \alpha(\psi,\tor, \theta = 0), \theta = 0) = A(\psi,\alpha(\psi,\tor, \theta =2 \upi N ), \theta = 2 \upi N).
\end{equation}
Here, $N$ is an integer indicating the number of poloidal turns that is chosen as an input parameter for simulation.

To illustrate the difficulty of applying this boundary condition in the general case of arbitrary  $q$ profile, we first consider the local case where $\psi$ is used as the radial coordinate. Then, with $\alpha = \tor - q(\psi) \vartheta$,~\eqref{eqn:parBC} can be expressed in Fourier space as
\begin{equation}
   \sum_{k_\psi, k_\alpha} A_\bb{k}(\theta = 0) \rme^{\imag k_\psi (\psi-\psi_0) + \imag k_\alpha (\alpha - \alpha_0)} = \sum_{k_\psi, k_\alpha} A_\bb{k}(\theta= 2\upi)\rme^{\imag k_\psi (\psi-\psi_0) + \imag k_\alpha (\alpha - \alpha_0)  - 2\upi N \imag k_\alpha [q_0 +  (\od q / \od \psi)(\psi-\psi_0) + (\od^2 q/ \od \psi^2) (\psi-\psi_0)^2/2 + \ldots]}.
\end{equation}
In the flux-tube limit, only the first two terms in the Taylor expansion of $q(\psi) \approx q_0 +  \od q/ \od \psi (\psi-\psi_0)$ are kept, and so the parallel boundary condition amounts to matching a quantity at either end of the $\vartheta$ domain at different radial wavenumber $k_\psi$,
\begin{equation}\label{eqn:twistshift_psi}
A_{k_\psi, k_\alpha}(\theta = 0) = C_k A_{k_\psi + \upDelta k_\psi,  k_\alpha}(\theta= 2\upi), 
\end{equation}
where $C_k = \exp(-2\upi N \imag k_\alpha  q_0  ) $ and $\upDelta k_\psi = 2\upi N k_\alpha (\od q / \od \psi)$. Equation~\eqref{eqn:twistshift_psi} allows for an efficient and straightforward implicit treatment of the parallel streaming term, which otherwise can set a stringent constraint on the simulation time step.  In practice, as the phase-shift due to $C_k$ is of order $\rho_\ast^{-1}$, a $\rho_\ast$-small adjustment in the radial position of the simulation domain allows for $C_k = 1$; this is equivalent to adjusting the positions of the mode rational surfaces that are introduced by the parallel boundary condition~\cite{Ball_prl,Ajay2020}.  

One must keep more terms in the Taylor expansion of $q(\psi)$ for global simulations, and as a result the parallel boundary condition cannot be expressed using a simple wavenumber shift. In general, the quantity would have to be inverse-Fourier transformed to $(x,k_y)$ space, have an $x$-dependent phase shift in the $y$ direction applied, and then Fourier transformed back into $(k_x,k_y)$ space, thus coupling all the $k_x$ modes at the $\theta$ boundaries. This complication can be avoided altogether if $q$ is chosen as the radial coordinate, rather than $\psi$. In this case,~\eqref{eqn:parBC} can be expressed in Fourier space as
\begin{equation}
   \sum_{k_q, k_\alpha} A_\bb{k}(\theta = 0) \rme^{\imag k_q (q-q_0) + \imag k_\alpha (\alpha - \alpha_0)} = \sum_{k_q, k_\alpha} A_\bb{k}(\theta= 2\upi)\rme^{\imag (k_q - 2 \upi N k_\alpha)(q-q_0) + \imag k_\alpha (\alpha - \alpha_0)  - 2\upi N \imag k_\alpha q_0 },
\end{equation}
and so now 
\begin{equation}\label{eqn:twistshift_q}
A_{k_q, k_\alpha}(\theta = 0) = C_k A_{k_q + \upDelta k_q, k_\alpha}(\theta= 2\upi), 
\end{equation}
where $C_k = \exp(-2\upi N \imag k_\alpha  q_0  ) $ and $\upDelta k_q = 2\upi N k_\alpha $. This, however, comes at the cost of only allowing global simulations of systems with a monotonic $q$ profile. We thus choose $q$ as the radial coordinate for global \texttt{stella}, though alternative formulations of the parallel boundary condition will be investigated in future work.

For global simulations, the radial coordinate $x$ (equivalently, $q$) can be related to the physical location $r$ using
\begin{align}
q-q_0 =\tilde{x}\rho_\ast\frac{aB_\mr{ref}}{\psi'} = q'\rarg  (\rcoord - \rcoord_0) + \frac{q''\rarg}{2} (\rcoord-\rcoord_0)^2 + \cdots
\end{align}
where $\tilde{x}\doteq x/\rho_\mr{ref}$ and~\eqref{eqn:xcoord_glob} has been used for the first equality. 
Defining $\upDelta \rcoord = \rcoord - \rcoord_0$ and keeping terms up to order $\upDelta \rcoord^2$, the physical location in terms of $x$ is given by the positive root of the quadratic equation,
\begin{align}\label{eqn:grid_quad}
\upDelta \rcoord = \frac{q'}{q''}\left[-1  + \left(1 + 2 \tilde{x} \rho_\ast \frac{q''}{q'^2}\frac{aB_\mr{ref}}{ \psi'} \right)^{1/2}\right].
\end{align}
These results are readily generalized if $\psi$ is used as the radial coordinate instead.

The extent of the radial domain in $q$ must be chosen to ensure $\rcoord/a \in (0,1)$. Note that for $q'' \ne 0$, a grid centered in $q$ around $q_0$ will not be centered in $\rcoord$ around $\rcoord_0$. By default, global \texttt{stella} constructs the radial grid to be centered in $\rcoord$, which leads to the two constraint equations 
\begin{align*}
  \frac{q''}{2q'} \upDelta \rcoord_+^2  - \upDelta \rcoord_+ - \tilde{x}_-\rho_\ast \frac{aB_\mr{ref}}{q'\psi'}  &=  0, \\
  \frac{q''}{2q'} \upDelta \rcoord_+^2  + \upDelta \rcoord_+ -  \tilde{x}_+\rho_\ast \frac{aB_\mr{ref}}{q'\psi'}  &=  0, 
\end{align*}
where $\tilde{x}_\pm$ are the left and rightmost values of the $\tilde{x}$ grid, $\tilde{x}_+ = \tilde{x}_- + \tilde{\ell}_x $,  $\tilde{\ell}_x\doteq \ell_x/\rho_\mr{ref}$, and $\upDelta r_+ = \rcoord_+ - \rcoord_0$ where $\rcoord_+$ is the rightmost point of the radial grid. Solving this system for $\upDelta \rcoord$ and $\tilde{x}_-$ gives
\begin{subequations}
\begin{align}\label{drho}
  \upDelta \rcoord_+ &= \frac{ \tilde{\ell}_x\rho_\ast}{2} \frac{aB_\mr{ref}}{q'\psi'}, \\
  \tilde{x}_- &= -\frac{\tilde{\ell}_x}{2} \left( 1- \frac{ \tilde{\ell}_x\rho_\ast}{4} \frac{q''}{q'^2}\frac{aB_\mr{ref}}{\psi'}\right).
\end{align}
\end{subequations}
Thus, the physical radial box size $\ell_\rcoord = 2\upDelta r_+ =(aB_\mr{ref}/q'\psi') \tilde{\ell}_x\rho_\ast$ is determined by the input parameters $\rho_\ast$, $\tilde{\ell}_x$ and $q'$, along with $\psi'$ calculated from the Miller geometry (\ref{app:mag_geo}).
If $q'' = 0$, then $\tilde{x}_- =- \tilde{\ell}_x/2$ and the grid becomes centered in $x$ (or $q$) as well.

\subsection{Radial boundary conditions}\label{sec:radBC}

\begin{figure}
    \centering
    \includegraphics[width=0.8\textwidth]{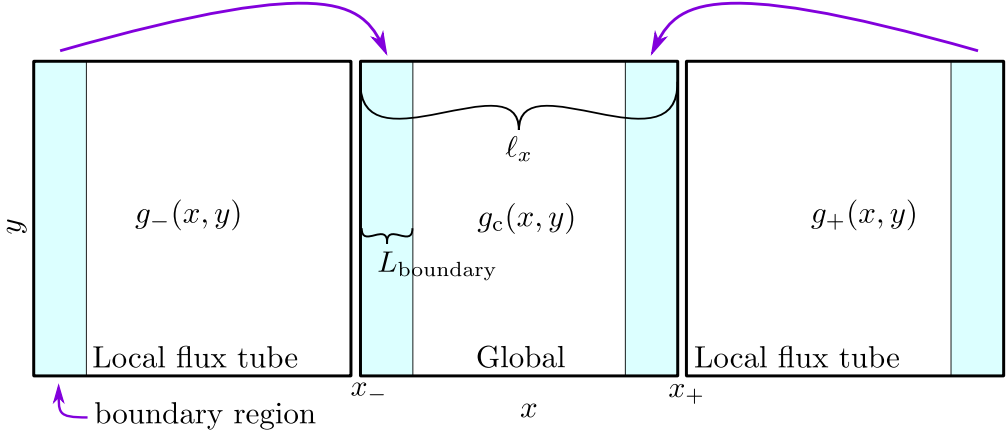}
    \caption{Illustration of the novel radial boundary condition which employs multiple flux-tube simulations.  The simulations on the left and right, which are local simulations, are performed concurrently to and indepedent of the central `global' simulation. At every timestep, information of the distribution function from  the side domains is directly copied into analogous regions in the central domain. This transfer of information is done in specified boundary regions. }
    \label{fig:radial_BC}
\end{figure}

The global version of \texttt{stella} employs a novel set of radial boundary conditions which feeds information generated from independent local flux-tube simulations into a central `global' simulation. The aim of these boundary conditions is two-fold: firstly, we replace the Dirichlet boundary condition typically used in global simulations with a more physically motivated forcing. By doing so, particle, momentum, and heat fluxes, which are self-consistently generated by local simulation, can be specified at the boundaries in a statistical sense. The second aim of our novel boundary conditions is that by feeding in `good' information at the radial boundaries, the Gibbs phenomena that typically arise near a discontinuity when using a Fourier spectral approach is largely mitigated. 

The novel radial boundary condition, which is illustrated in figure~\ref{fig:radial_BC}, works as follows: a boundary region is specified for each flux-tube domain with a given number of radial collocation points $N_\mr{boundary}$ and radial width $\tilde{L}_\mr{boundary}$; the domains of the local simulations each contain a single instance of this boundary, while the central domain contains a boundary on each radial end. Then, at every timestep (or, alternatively, every Runge-Kutta explicit and implicit substep), information from the distribution functions in the left and right domains ($g_{s, \mr{L}}$ and $g_{s,\mr{R}}$, respectively) is copied directly into the boundary region of the distribution function of the central domain:
\begin{subequations}\label{eqn:radial_BC}
\begin{align}
\tilde{g}_{s}(\tilde{x} +  \tilde{x}_-) &= g_{s,\mr{L}}(\tilde{x} +  \tilde{x}_{-,\mr{L}}), \\
\tilde{g}_{s}(\tilde{x}+  \tilde{x}_+  - \tilde{L}_\mr{boundary}) &= g_{s,\mr{R}}(\tilde{x}+  \tilde{x}_{+,\mr{R}}  - \tilde{L}_\mr{boundary}), 
\end{align}
\end{subequations}
for $0 < \tilde{x} < \tilde{L}_\mr{boundary}$. The locations of the boundary regions here are chosen to be consistent with the parallel boundary condition: while local simulations employing the parallel boundary condition are periodic in the radial direction, they are not homogeneous due to the appearance of mode rational surfaces at distinct radial locations. We stress here that these radial locations remain equally spaced due to our choice of $q$ as the radial coordinate, regardless of the profile of the magnetic safety factor. The placement of the boundary regions outlined in~\eqref{eqn:radial_BC} and illustrated in figure~\ref{fig:radial_BC} ensures that these mode rational surfaces line up correctly when the parallel-boundary phase shift $C_k$ is chosen to be consistent in every domain (see \S \ref{sec:coords}). Intuitively, the size of the $\tilde{L}_\mr{boundary}$ must be larger than the largest Larmor radius contained in the simulation in order for information not to leak through the edge of the radial simulation domain. Numerical tests show that physical observables such as the heat flux converge once the boundary layer size meets this criteria, and that numerical instabilities may appear if the region is too small.

The two auxiliary local simulations are seeded with different initial conditions than the central domain and all three are run concurrently with the same timestep (which may be adjusted during the simulation based on the CFL constraint). Currently, in order to avoid interpolation when mapping onto the central domain, the left and right domains use the same grid resolution and spacing in the parallel and binormal coordinates, and also share the same velocity space grid. While the grid spacing in the radial direction must also match, the left and right simulation domain are free to have different radial extents. While this set-up effectively triples the computational cost of the equivalent local simulation, it does so in an embarrassingly parallel way; a fact of which the parallelism of~\texttt{stella} readily takes advantage.


\section{Numerical Implementation \label{sec:numerical_imp}} 

\subsection{Simulation equations \label{sec:sim_equations}}

Global \texttt{stella} uses a pseudospectral approach to integrate the gyrokinetic equation. Here, the linear terms with constant coefficients are evaluated in Fourier space, while nonlinear terms and terms with radial profile variation are evaluated in real space. The Fourier transformed gyrokinetic equation, including the radial variation terms resulting from the Taylor expansion outlined in \S \ref{sec:subsidiary}, is


\begin{align}\label{eqn:GKE_FT}
\frac{\partial \tilde{g}_{\bb{k},s}}{\partial t} &+ \imag k_y\tilde{\Omega}_\tor  \tilde{g}_{\bb{k},s} + \frac{v_{\mr{th}s}\rarg}{v_{\mr{th,ref}}} \tilde{v}_\parallel {\eb \bcdot \tilde{\grad} \tilde{z}} \Bigg(\frac{\partial \tilde{g}_{\bb{k},s}}{\partial \tilde{z}}  + {\frac{Z_s}{\tilde{T}_s\rarg} } \frac{\partial J_0(a_{\bb{k},s}\rarg) \tilde{\varphi}_{\bb{k}} }{\partial \tilde{z}}\rme^{-v^2\rarg/v^2_{\mr{th}s}\rarg} \Bigg) - \frac{v_{\mr{th}s}\rarg}{v_{\mr{th,ref}}} \tilde{\mu}_s{\eb \bcdot \tilde{\grad} \tilde{B}} \frac{\partial \tilde{g}_{\bb{k},s}}{\partial \tilde{v}_\parallel} \nonumber\\
&+ \imag \tilde{\omega}_{\mr{D},\bb{k},s}\left( \tilde{g}_{\bb{k},s} +{\frac{Z_s}{\tilde{T}_s\rarg}} J_0(a_{\bb{k},s}\rarg) \tilde{\varphi}_{\bb{k}}\rme^{-v^2\rarg/v^2_{\mr{th}s}\rarg} \right) 
+  (\imag\tilde{\omega}_{\ast,\bb{k},s} + \imag\tilde{\omega}_{\tor,\bb{k},s})J_0(a_{\bb{k},s})\tilde{\varphi}_\bb{k} + \mathcal{N}_{\bb{k},s} \nonumber\\
= &- \frac{v_{\mr{th}s}\rarg}{v_{\mr{th,ref}}} \tilde{v}_\parallel \left({\eb \bcdot \tilde{\grad} \tilde{z}}\right)' \spaceOperator \Bigg(\frac{\partial \tilde{g}_{\bb{k},s}}{\partial \tilde{z}}  + {\frac{Z_s}{\tilde{T}_s\rarg} } \frac{\partial J_0(a_{\bb{k},s}\rarg) \tilde{\varphi}_{\bb{k}} }{\partial \tilde{z}}\rme^{-v^2\rarg/v^2_{\mr{th}s}\rarg} \Bigg) + \frac{v_{\mr{th}s}\rarg}{v_{\mr{th,ref}}} \tilde{\mu}_s\left(\eb \bcdot \tilde{\grad} \tilde{B}\right)' \spaceOperator \frac{\partial \tilde{g}_{\bb{k},s}}{\partial \tilde{v}_\parallel} \nonumber \\
& - \frac{v_{\mr{th}s}\rarg}{v_{\mr{th,ref}}} \tilde{v}_\parallel {\eb \bcdot \tilde{\grad} \tilde{z}} \spaceOperator \Bigg[  \left(\frac{F_{s}(\rcoord)}{\tilde{T}_s}\right)'\frac{\partial J_0(a_{\bb{k},s}\rarg) \tilde{\varphi}_{\bb{k}} }{\partial \tilde{z}} \frac{Z_s\rme^{-v^2\rarg/v^2_{\mr{th}s}\rarg}}{F_s(\rcoord_0)} + {\frac{Z_s}{\tilde{T}_s\rarg} } \frac{\partial J'_0(a_{\bb{k},s}) \tilde{\varphi}_{\bb{k}} }{\partial \tilde{z}}\rme^{-v^2\rarg/v^2_{\mr{th}s}\rarg} \Bigg] \nonumber\\
&- \spaceOperator \left[\left( \imag\tilde{\omega}_{\mr{D},\bb{k},s}\right)'\left( \tilde{g}_{\bb{k},s} +{\frac{Z_s}{\tilde{T}_s\rarg}} J_0(a_{\bb{k},s}\rarg) \tilde{\varphi}_{\bb{k}}\rme^{-v^2\rarg/v^2_{\mr{th}s}\rarg} \right)\right]
- \spaceOperator\left[\left( ( \imag \tilde{\omega}_{\ast,\bb{k},s} +  \imag \tilde{\omega}_{\tor,\bb{k},s})J_0(a_{\bb{k},s})\right)'\tilde{\varphi}_\bb{k} \right]\nonumber \\
&- \spaceOperator\Bigg[ \left(\frac{F_{s}(\rcoord)}{\tilde{T}_s}\right)' \imag \tilde{\omega}_{\mr{D},\bb{k},s} J_0(a_{\bb{k},s}\rarg) \tilde{\varphi}_{\bb{k}}  \frac{Z_s\rme^{-v^2\rarg/v^2_{\mr{th}s}\rarg}}{F_s(\rcoord_0)}  \Bigg] 
- \imag \tilde{\omega}_{\mr{D},\bb{k},s}\spaceOperator\Bigg( {\frac{Z_s}{\tilde{T}_s\rarg} }  J'_0(a_{\bb{k},s}) \tilde{\varphi}_{\bb{k}} \rme^{-v^2\rarg/v^2_{\mr{th}s}\rarg} \Bigg) + \left(\mathcal{N}_{\bb{k},s}\right)', 
\end{align}
where all quantities that depend on $\rcoord$ have been evaluated at $\rcoord_0$. 
Here, $a_{\bb{k},s} =(\tilde{m}_s \tilde{T}_s/Z_s^2)^{1/2} \tilde{k}_\perp \tilde{v}_\perp / \tilde{B}$, $ \tilde{k}_\perp^2 = k_x^2\rho_\mr{ref}^2 |\grad x|^2 + 2  k_x k_y \rho_\mr{ref}^2(\grad x \bcdot \grad y) + k^2_y\rho_\mr{ref}^2 |\tilde{\grad} y|^2$, 
\begin{subequations}
\begin{align}
    \tilde{\omega}_{\mr{D},\bb{k},s} &= \frac{a\tilde{T}_s(\rcoord_0)}{Z_s \tilde{B}(\rcoord)}\left(\tilde{v}_\parallel^2 \eb \btimes \bb{\kappa} + \tilde{\mu}_s \eb \btimes \grad \tilde{B}\right)\bcdot\left(k_y \rho_\mr{ref} \grad y + k_x \rho_\mr{ref}\grad x\right),\\
    \tilde{\omega}_{\ast,\bb{k},s} &= \frac{k_y \rho_\mr{ref}}{2} \frac{aB_\mr{ref}}{\psi'} \frac{\od y}{\od \alpha}\frac{\rme^{-v^2\rarg/v^2_{\mr{th}s}\rarg}}{F_s(\rcoord_0)}\left.\frac{\od  F_s(\rcoord)}{\od \rcoord}\right|_E,\\
    \tilde{\omega}_{\tor,\bb{k},s} &=  -k_y\rho_\mr{ref}\sqrt{\frac{\tilde{m}_s}{\tilde{T}_s}} \frac{q a}{r}\frac{\tilde{I}}{\tilde{B}} \tilde{v}_\parallel \rme^{-v^2\rarg/v^2_{\mr{th}s}\rarg} \tilde{\gamma}_E,
\end{align}
\end{subequations}
$\tilde{\gamma}_E = (\rcoord/q)(\od \Omega_\zeta /\od \rcoord) (a/v_\mr{th,ref})$,
and the nonlinear term for when $q$ is the radial coordinate is given by
\begin{equation}
\mathcal{N}_{\bb{k},s} = \frac{B_\mr{ref}}{2}\frac{\od y }{\od \alpha}\frac{\od x}{\od q}\frac{q'}{\psi'}\fourier\left[\fourier^{-1}\left(\imag k_y \rho_\mr{ref}J_0(a_{\bb{k},s})\tilde{\varphi}_\bb{k}\right)\fourier^{-1}\left(\imag k_x \rho_\mr{ref}\tilde{g}_{\bb{k},s}\right) - \fourier^{-1}\left(\imag k_x \rho_\mr{ref}J_0(a_{\bb{k},s})\tilde{\varphi}_\bb{k}\right)\fourier^{-1}\left(\imag k_y \rho_\mr{ref}\tilde{g}_{\bb{k},s}\right)\right],
\end{equation}
where $\fourier$ and $\fourier^{-1}$ denote the forward and inverse Fourier transforms, respectively. 

The terms dealing with radial profile variation are denoted using a prime; here,
\begin{subequations}
\begin{align}
\left(\frac{F_{s}(\rcoord)}{\tilde{T}_s}\right)' &=  \frac{F_{s}}{\tilde{T}_s} \left[ \frac{n'_s}{n_s} + \frac{T'_s}{T_s}\left(\frac{E}{T_s} - \frac{5}{2}\right)- \frac{\mu_s B'}{T_s}\right], \\ 
\left(\frac{1}{\psi' } \left. \frac{\od  F_{s}(\rcoord)}{\od \rcoord} \right|_E \right)' &=  \frac{1}{\psi'} \bigg\{ \frac{n''_s}{n_s} - \frac{n'^2_s}{n_s^2} + \left(\frac{T''_s}{T_s}  - \frac{T'^2_s}{T_s^2}\right)\left(\frac{ E}{T_s}-\frac{3}{2}\right)- \frac{T'^2_s}{T_s^2}\frac{E}{T_s}  + \frac{T'_s}{T_s} \frac{\mu_s B'}{T_s} \nonumber \\
&\quad+ \left[\frac{n'_s}{n_s} + \frac{T'_s }{T_s}\left(\frac{ E}{T_s}-\frac{3}{2}\right) \right]\left[\frac{n'_s}{n_s}+ \frac{T'_s}{T_s}\left(\frac{ E}{T_s}-\frac{3}{2}\right) - \frac{\mu_s B'}{T_s}- \frac{\psi''}{\psi'}\right]\bigg\}
F_{s},  \\
\left(\mathcal{N}_{\bb{k},s}\right)' &=\spaceOperator \left(  \frac{q''}{q'} - \frac{\psi''}{\psi'}\right) \mathcal{N}_{\bb{k},s} + \frac{B_\mr{ref}}{2}\frac{\od y }{\od \alpha}\frac{\od x}{\od q}\frac{q'}{\psi'}\fourier\bigg[\fourier^{-1}\left(\imag k_y  \rho_\mr{ref} \spaceOperator J'_0(a_{\bb{k},s})\tilde{\varphi}_\bb{k}\right)\fourier^{-1}\left(\imag k_x \rho_\mr{ref}\tilde{g}_{\bb{k},s}\right) \nonumber\\
&\qquad - \fourier^{-1}\left(\imag k_x  \rho_\mr{ref} \spaceOperator J'_0(a_{\bb{k},s})\tilde{\varphi}_\bb{k}\right)\fourier^{-1}\left(\imag k_y \rho_\mr{ref}\tilde{g}_{\bb{k},s}\right)\bigg],\\
J'_0(a_{\bb{k},s}) &= - J_1(a_{\bb{k},s})a_{\bb{k},s}\left(\frac{({k}_\perp^2)'}{{k}_\perp^2} - \frac{{B}'}{{B}}\right).
\end{align}
\end{subequations}
In these equations, tildes have been suppressed for fractional terms that are dimensionless. 
The other radial profile terms dealing with the magnetic geometry (see \S \ref{sec:var_outline}) are computed in  \ref{app:mag_geo}. The radial operator $\spaceOperator$ applies the $\rcoord$ profile to all the terms appearing to the right of it, and is defined as
\begin{equation}\label{eqn:spaceOperator}
    \spaceOperator (\cdots) \doteq \fourier \left\{ (\rcoord_\mr{clamped} - \rcoord_0) \fourier^{-1}\left\{ \cdots\right\}\right\},
\end{equation}
where
\begin{equation}
\rcoord_\mr{clamped} \doteq \left\{ \begin{matrix}
\rcoord(\tilde{x}_- + \tilde{L}_\mr{boundary}) & \tilde{x}_- \le \tilde{x} \le   \tilde{x}_- + \tilde{L}_\mr{boundary},\\
\rcoord(\tilde{x}) &  \tilde{x}_- + \tilde{L}_\mr{boundary} \le \tilde{x} \le   \tilde{x}_+ - \tilde{L}_\mr{boundary},\\
\rcoord(\tilde{x}_+ - \tilde{L}_\mr{boundary}) & \tilde{x}_+ - \tilde{L}_\mr{boundary} \le \tilde{x} \le   \tilde{x}_+.
\end{matrix} \right.
\end{equation}
We use this definition of $\rcoord_\mr{clamped}$ in order to render the evolution of $g_s$ within the radial boundary region as consistent as possible between the three simulation domains. Global \texttt{stella} also includes an option to employ the radial boundary conditions suggested by Candy et al.~\citep{Candy_globshear,Candy_globloc} by using a triangle waveform in the definition of~\eqref{eqn:spaceOperator}. Additionally, both approaches can be combined in order to further mitigate Gibbs phenomena in the higher radial derivatives of the distribution function and electrostatic potential. This is done by placing one of the boundary regions of the central domain into the middle of the box.

Adding global effects as a next-order correction in a Taylor expansion provides two advantages over using arbitrary profile variation: Firstly, as all the global terms utilize the same spatial operator $\spaceOperator$, it greatly reduces the number of Fourier transforms needed to implement the terms numerically. Secondly, as only quadratic variations of the kinetic and magnetic geometry profiles are added, there are fewer free parameters needed to specify the physical system, and so the problem of 'flux-matching' global simulations to experimental results is somewhat simplified. Future versions of global~\texttt{stella} will aim to allow arbitrary profile variation in both the kinetic and magnetic profiles, though the planned inclusion of electromagnetic effects will make this a more challenging task.

 The Fourier-transformed quasineutrality equation is 
\begin{align}
 \sum_s  Z_s & \tilde{n}_s(\rcoord_0)  \frac{2}{\uppi^{1/2}} \int \od \tilde{v}_\parallel \int \od \tilde{\mu}_s \,\tilde{B}\left(J_0(a_{\bb{k},s}\rarg) \tilde{g}_{\bb{k},s} + \frac{Z_s }{\tilde{T}_s(\rcoord_0)} \rme^{-v^2\rarg/v^2_{\mr{th}s}\rarg}\left(J^2_0(a_{\bb{k},s}) -1\right)\tilde{\varphi}_{\bb{k}}\right) \nonumber \\
 = &-\spaceOperator\sum_s  Z_s \tilde{n}_s(\rcoord_0)  \frac{2}{\uppi^{1/2}} \int \od \tilde{v}_\parallel \int \od \tilde{\mu}_s \tilde{B}\Bigg\{\left(J'_0(a_{\bb{k},s}) + \frac{B'}{B}J_0(a_{\bb{k},s}) \right) \tilde{g}_{\bb{k},s}  \nonumber
 \\ & + \frac{Z_s }{\tilde{T}_s(\rcoord_0)} \rme^{-v^2\rarg/v^2_{\mr{th}s}\rarg}\left(J^2_0(a_{\bb{k},s}) -1\right) \left[ \frac{n'_s}{n_s} + \frac{T'_s}{T_s}\left(\frac{E}{T_s} - \frac{5}{2}\right) + \frac{B'}{B}\left(1 - 2 \tilde{\mu}_s \tilde{B}\right) - \frac{2J_0(a_{\bb{k},s})J'_0(a_{\bb{k},s})}{1-J^2_0(a_{\bb{k},s})}\right]\tilde{\varphi}_{\bb{k}}\Bigg\}.
\end{align}
Solving quasineutrality requires special considerations, which are given in \S \ref{sec:quasi_solve}.

 Some comments on the Fourier transforms in the spatial operator $\spaceOperator$ are in order. The inclusion of terms with a linear profile $\rcoord - \rcoord_0$ introduces the effect of profile shearing, which results in the advection of the radial wavenumber of a mode in Fourier space~\citep{Candy_globloc}.  In the case of a positive advection speed in $k_x$, a mode's radial wavenumber may reach the maximal radial wavenumber  resolved in the simulation, $k_{x,\rm{max}}$. If no other steps are taken, it will then unphysically wrap around to the opposite extreme at $k_{x,\rm{min}} = -k_{x,\mr{max}}$. (For negative advection speed in $k_x$, wrap-around from $k_{x,\rm{min}}$ to $k_{x,\rm{max}}$ will instead occur.) This is related to the phenomena of aliasing which could occur  when the distribution function $g_{\bb{k},s}$ and electrostatic potential $\varphi_\bb{k}$ are inverse Fourier transformed to real space in order to calculate the nonlinear term $\mathcal{N}_{\bb{k},s}$. To avoid aliasing by the nonlinear term, some form of dealiasing is performed, the conventional method being the `2/3rds approach' ~\citep{Orszag_aliasing}, where $g_{\bb{k},s}$ and $\varphi_\bb{k}$ are inverse Fourier transformed onto a grid with 50\% more collocation points in each Fourier transformed dimension than the number of corresponding Fourier modes. This approach to dealiasing for the nonlinear term can be justified as a spectral cut-off of the distribution function and electrostatic potential that can be enforced with sufficient (hyper)viscosity, and thus their truncation when transforming back to Fourier space should not have any significant effect. However, this approach is unsuitable for the operator  $\spaceOperator$, as truncation of the discontinuous function $\rcoord - \rcoord_0$ introduces Gibbs phenomena which results in numerical artefacts when convolved with another term. Additionally, in our experience dealiasing by spectral cut-off does not prevent wrap-around.
 
 The Fourier transforms that appear in the spatial operator $\spaceOperator$  are thus performed without any additional padding, relying only on hyper-dissipation (\S \ref{sec:hyperdissipation}) to prevent radial wavenumber wrap-around due to profile shearing. This approach can be justified by the smallness of profile shearing due to $\spaceOperator$: while $(\rcoord - \rcoord_0)/a$ is of order $\Delta$, profile shearing depends on its derivative $\od\, (\rcoord - \rcoord_0)/\od x$, which is formally of order $\rho_\ast \ll \Delta$. The hyper-dissipative damping rate at the radial wavenumber grid boundaries can then be made to be faster than the rate at which a radial mode advects across of radial wavenumber cell of width $\upDelta k_x = 2 \upi/\ell_x$.

\subsection{Integration scheme \label{sec:int_scheme}}

The \texttt{stella} code, either in local flux-tube or global operation, employs an operator-split time integration scheme that alleviates the timestep constraint caused by the fast parallel electron dynamics. In the local version, this scheme comprises three main steps~\citep{Barnes_stella},
\begin{equation}
\frac{\partial \tilde{g}_{\bb{k},s}}{\partial t} =\left(\frac{\partial \tilde{g}_{\bb{k},s}}{\partial t}\right)_1 +\left(\frac{\partial \tilde{g}_{\bb{k},s}}{\partial t}\right)_2
 + \left(\frac{\partial \tilde{g}_{\bb{k},s}}{\partial t}\right)_3,
\end{equation}
where
\begin{subequations}\label{eqn:timestep}
\begin{align}
\left(\frac{\partial \tilde{g}_{\bb{k},s}}{\partial t}\right)_1 &= -\imag \omega_{\mr{D},\bb{k},s}\left(\tilde{g}_{\bb{k},s} + J_0(a_{\bb{k},s}) \tilde{\varphi}_\bb{k}\frac{Z_s}{\tilde{T}_s} \frac{F_s(\rcoord)}{F_s(\rcoord_0)}\rme^{-{v^2(\rcoord_0)}/v_{\mr{th}s}^2(\rcoord_0)}  \right)- \imag \omega_{\ast,\bb{k},s}J_0(a_{\bb{k},s})\tilde{\varphi}_\bb{k} - \mathcal{N}_\bb{k}, \label{eqn:timestep_exp}\\ 
\left(\frac{\partial \tilde{g}_{\bb{k},s}}{\partial t}\right)_2 &= \frac{v_{\mr{th},s}(\rcoord_0)}{v_{\mr{th,ref}}}\tilde{\mu}_s \eb\bcdot \tilde{\grad} \tilde{B} \frac{\partial \tilde{g}_{\bb{k},s}}{\partial \tilde{v}_\parallel},\label{eqn:timestep_mir}\\ 
\left(\frac{\partial \tilde{g}_{\bb{k},s}}{\partial t}\right)_3 &= -\frac{v_{\mr{th},s}(\rcoord_0)}{v_{\mr{th,ref}}}v_\parallel \eb\bcdot \tilde{\grad} z \left( \frac{\partial \tilde{g}_{\bb{k},s}}{\partial z} + \frac{\partial J_0(a_{\bb{k},s}) \tilde{\varphi}_\bb{k}}{\partial z}\frac{Z_s}{\tilde{T}_s} \frac{F_s(\rcoord)}{F_s(\rcoord_0)}\rme^{-{v^2(\rcoord_0)}/v_{\mr{th}s}^2(\rcoord_0)}  \right).\label{eqn:timestep_prl}
\end{align}
\end{subequations}
The first step, which includes the magnetic and $\bb{E} \btimes \bb{B}$ drifts, is an explicit step which utilizes a strong stability preserving, third-order Runge-Kutta method. The next two steps, which respectively perform the parallel acceleration and streaming, are done implicitly. Additional implicit steps, such as those performing equilibrium flow shear or applying dissipative operators, can also be included in the time integration scheme.

Global \texttt{stella} incorporates global effects by computing the terms on the right-hand-side of \eqref{eqn:GKE_FT} that account for the radial variation of pressure and magnetic geometry, and including them in the explicit step given by~\eqref{eqn:timestep_exp}. The explicit treatment of the radial corrections is not expected to place a limit on the time step, provided $\Delta$ is not too large; indeed, we have found this to be the case for the ion scale simulations that we have performed. { Finally, while the parallel streaming term~\eqref{eqn:timestep_prl} can in principle be calculated implicitly for global~\texttt{stella}, for reasons detailed in \S \ref{sec:quasi_solve} this term is calculated explicitly by default, as well as for all the numerical benchmarks performed in this paper; future versions of \texttt{stella} will include implicit global algorithms for the parallel dynamics. }

\subsection{Quasineutrality \label{sec:quasi_solve}}

At every integration sub-step that involves the electrostatic potential, the quasineutrality equation~\eqref{eqn:quasi_norm} must be used to recompute $\tilde{\varphi}_\bb{k}$ using the updated value of $\tilde{g}_{\bb{k},s}$. This involves performing a velocity-space integration of the distribution function, and then an inversion of the spectral double-gyro-average operator
\begin{equation}
    Q= \sum_s \int \od^3 v \,   (Z_s^2 e/T_s) F_s \left(1-J_0^2(a_{\bb{k},s})\right).
\end{equation}
The latter step is trivial in local flux-tube simulations as the operator is a diagonal operator in $\bb{k}$-$\theta$ space. In global operation, the quasineutrality equation including global profile variation can be expressed as 
\begin{equation}\label{eqn:quasi_rad}
    \Theta \tilde{\varphi}_\bb{k} +\spaceOperator(\Theta'\tilde{\varphi}_\bb{k}) = \frac{2}{\upi^{1/2}}\sum_s Z_s \int \od \tilde{v}_\parallel \int \od \tilde{\mu}_s \left\{ \tilde{B}(\rcoord_0) J_0(a_{\bb{k},s}\rarg)\tilde{g}_{\bb{k},s} + \spaceOperator \left[ \left(\tilde{B}' J_0(a_{\bb{k},s}) + \tilde{B} J'_0(a_{\bb{k},s})\right)_{\rcoord = \rcoord_0}\tilde{g}_{\bb{k},s}\right]\right\},
\end{equation}
where 
\begin{subequations}
\begin{align}
\Theta &=\sum_s \frac{2}{\pi^{1/2}}\int \od \tilde{v}_\parallel \od \tilde{\mu}_s \, \tilde{B}(\rcoord_0) Z_s^2 \frac{\tilde{n}_s(\rcoord_0)}{\tilde{T}_s(\rcoord_0)}\left(1-J_0^2(a_{\bb{k},s}\rarg) \right)\, \rme^{-v^2\rarg/v^2_{\mr{th}s}\rarg},\\
  \Theta' &= \sum_s \frac{2}{\pi^{1/2}}\int \od \tilde{v}_\parallel \od \tilde{\mu}_s \, \tilde{B}(\rcoord_0) Z_s^2 \frac{\tilde{n}_s(\rcoord_0)}{\tilde{T}_s(\rcoord_0)}\left(1-J_0^2(a_{\bb{k},s}\rarg) \right)\, \rme^{-v^2\rarg/v^2_{\mr{th}s}\rarg} \times \nonumber
 \\ & \hspace{6cm} \left[ \frac{n'_s}{n_s} + \frac{T'_s}{T_s} \left(\frac{E}{T_s} - \frac{5}{2}\right) + \frac{B'}{B}\left(1 - 2 \tilde{\mu}_s \tilde{B}\right) - \frac{2J_0(a_{\bb{k},s})J'_0(a_{\bb{k},s})}{1-J^2_0(a_{\bb{k},s})}\right]_{\rcoord=\rcoord_0}.
\end{align}
\end{subequations}
Solving the quasineutrality equation~\eqref{eqn:quasi_rad} requires two steps: first the velocity integration of $\tilde{g}_{\bb{k},s}$ is carried out (which includes taking the radial variation of $B$, $k_\perp$ and $v_\perp$ into account). The second step is then solving for $\varphi$. However, the Fourier transforms that appear on the left-hand-side of~\eqref{eqn:quasi_rad} add the complication that every radial mode couples together, and so global \texttt{stella} offers the user two options in solving for $\varphi$: a perturbative approach and an exact approach. 

In the former, the electrostatic potential is decomposed into two pieces $\tilde{\varphi}_\bb{k} = \varphi_0 + \varphi_1$, where  
\begin{align}
    \varphi_0 &= \frac{1}{\Theta}\frac{2}{\upi^{1/2}}\sum_s Z_s \int \od \tilde{v}_\parallel \int \od \tilde{\mu}_s \tilde{B}(\rcoord_0) J_0(k_\perp(\rcoord_0)v_\perp(\rcoord_0)/\Omega_s(\rcoord_0))\tilde{g}_{\bb{k},s}, \\
    \varphi_1 &=  \frac{1}{\Theta}\spaceOperator\left[(2/\pi^{1/2})\sum_s Z_s \tilde{n}_s(\psi_0)\int \od \tilde{v}_\parallel \int \od \tilde{\mu}_s\, \tilde{g}_{\bb{k},s}\tilde{B}J_0(a_{s0})\left(\frac{J'_0}{J_0} +  \frac{\tilde{B}'}{B}- \frac{\Theta'}{\Theta }\right)_{\rcoord=\rcoord_0}\right].
\end{align}
Here, $\varphi_0$ is the electrostatic potential obtained in the $\rho_\ast$, $\Delta \rightarrow 0$ limit, while $\varphi_1$ is obtained by grouping the perturbative correction $\spaceOperator(\Theta'\varphi_0)$ together with the radial corrections of the gyroaverage and velocity integration of $\tilde{g}_{\bb{k},s}$. This new term, $\varphi_1$, is then included as an additional set of terms in the explicit step ~\eqref{eqn:timestep_exp}. As the variation of $\varphi_1$ is now included as a correction to the gyrokinetic equation, the zeroth-order portion of the parallel streaming term can remain implicit, and no further change in the response matrix approach is needed. The main drawback of the perturbative approach is sensitivity to any near-cancellations of $1- J_0^2(a_{\bb{k},s})$ at low wavenumbers, which is particularly problematic for $k_y=0$ zonal modes as information in higher wavenumber modes (for which $J_0^2(a_{\bb{k},s})-1$ is not small) may transfer to larger scales due to Fourier convolution with $\rcoord-\rcoord_0$, and so the $\Delta \ll 1$ expansion may break down even for moderately small $\Delta$.

Solving~\eqref{eqn:quasi_rad} exactly entails finding a solution to the linear equation
\begin{equation}\label{eqn:qn_full}
    \mathsfbi{Q}\bcdot \bb{\tilde{\varphi}}_{\bb{k}} = \bb{G},
\end{equation}
where $\mathsfbi{Q}\doteq  \Theta + \spaceOperator\Theta'$ is the complete quasineutrality operator, $\tilde{\bb{\varphi}}_{\bb{k}}$ is the column vector comprised of all $k_x$ modes of $\tilde{\varphi}_\bb{k}$ at a given $z$ location and binormal mode-number $k_y$,  and $\bb{G}$ is the right-hand-side of~\eqref{eqn:quasi_rad} expressed as a column vector. Solving~\eqref{eqn:qn_full} can be done efficiently through the use of an LU decomposition and back substitution~\citep{numerical_recipes}. Thus, $\varphi$ encompasses all the radial variation in the quasineutrality equation. (Radial variation of the gyroaveraging of $\varphi$ appearing in the gyrokinetic equation is still treated separately.) This approach unfortunately has the drawback of complicating the implicit solve of the parallel streaming term: the tridiagonal solve of~\eqref{eqn:timestep_prl} couples all modes connected by the parallel boundary condition, while the quasineutrality equation couples all modes radially, and so for a given binormal mode-number $k_y$, all radial and parallel grid points are coupled, resulting in a linear equation with a matrix of size $(N_xN_z)^2$, where $N_x$ is the number of radial modes and $N_z$ is the number of grid points in the parallel direction. Currently, when the full solve of quasineutrality is employed, global \texttt{stella} solves the parallel streaming term explicitly, though future versions of the code will include a modified response matrix algorithm that incorporates the exact quasineutrality equation.

Another complication that is introduced in the $\Delta$ expansion is the coupling of the $\varphi_{\bb{k} = \bb{0}}$ mode---which determines the parallel electric field $E_\parallel$---to other $\bb{k} \ne \bb{0}$ modes in the gyrokinetic equation. In the local limit, this mode does not enter the $\bb{E}\btimes \bb{B}$ nonlinearity nor does the non-linearity act on it; rather, it enters only in the parallel streaming term $v_\parallel\eb\bcdot \grad$, which is one-point in Fourier space in the local limit, and so the $\bb{k} =\bb{0}$ mode can be neglected.\footnote{Formally, this mode cannot be described in the flux-tube limit, as it does not obey the ordering $(k_\perp L)^{-1} \ll  1 $.} In the global limit, the $\bb{k} = \bb{0}$ mode of both the gyrokinetic and quasineutrality equations involve many $\bb{k}\ne \bb{0}$ modes, and so $\varphi_{\bb{k} = \bb{0}}$ should be solved self-consistently. 

The ease of determining $\varphi_{\bb{k}=\bb{0}}$ relies on how the species are treated; for cases with both kinetic ions and electrons, $\varphi_{\bb{k}=\bb{0}}$ is entirely absent in the quasineutality equation as $J_1 = J_0^2 - 1 = 0$ for $\bb{k} = \bb{0}$, and so $\varphi_{\bb{k}=0}$ must be determined from the parallel streaming term, using the $\bb{k} = \bb{0}$ quasineutrality equation at every $z$ location as its own solvability condition. This is most conveniently done when the parallel streaming is handled implicitly, and so the current version of global \texttt{stella}, which calculates this term explicitly, zeros out $\varphi_{\bb{k}=0}$. Future versions with the implicit parallel solve will aim to self-consistently evolve this mode.

In the case of adiabatic ions or electrons, $\varphi_{\bb{k}=\bb{0}}$ enters the quasineutrality equation through the Boltzmann response of the adiabatic species. For adiabatic ions with Boltzmann response $\delta n_\mr{i}/n_\mr{i} =(Z_\mr{i}e/T_\mr{i})\varphi$, including this mode is straightforward and equation~\eqref{eqn:qn_full} for $k_y = 0$ can be used as-is.
For adiabatic electrons, a modified Boltzmann response is used for the  electron density,
\begin{equation}\label{eqn:elec_boltz}
\frac{\delta n_\mr{e}}{n_\mr{e}} = \frac{q}{T_\mr{e}}\left(\varphi - \ba{\varphi}_\psi\right), 
\end{equation}
where the flux surface average $\ba{A}_\psi$ of a quantity $A$ is given by
\begin{equation}
    \ba{A}_\psi = \frac{\int \od y \int 
    \od z \, \mathcal{J} A }{\int \od y \int \od z \,\mathcal{J} } \approx  \frac{\int \od y \int 
    \od z \, \mathcal{J}\rarg A }{\int \od y \int \od z \,\mathcal{J}\rarg } + \frac{\rcoord_\mr{clamped} - 
    \rcoord_0}{\int \od y \int \od z \,\mathcal{J}\rarg}\int \od y \int \od z \, A \left(\mathcal{J}'\rarg - \mathcal{J}\rarg\frac{\int \od z \, \mathcal{J}'\rarg}{\int \od z \, \mathcal{J}\rarg}\right),
\end{equation}
where $\mathcal{J}$ and $\mathcal{J}'$ are given in~\ref{app:mag_geo}. 
Quasineutrality for the $k_y= 0$ mode then takes the form
\begin{equation}\label{eqn:quasi_rad_matrix}
   \mathsfbi{Q}\bcdot \bb{\tilde{\varphi}}_\bb{k} -  \bb{G} = -\mathsfbi{C}\bcdot ({\bb{\tilde{\varphi}}_\bb{k}}-\ba{\bb{\tilde{\varphi}}_\bb{k}}_\psi),
\end{equation}
where  $\mathsfbi{Q}$ now excludes the electron response and  $\mathsfbi{C} \doteq \tilde{n}_\mr{e}/\tilde{T}_\mr{e} + \spaceOperator (\tilde{n}_\mr{e}/\tilde{T}_\mr{e} )'$. The electron Boltzmann response~\eqref{eqn:elec_boltz} now imposes a solvability constraint on the quasineutrality equation
\begin{equation}\label{eqn:solvability}
  \langle \mathsfbi{Q}\bcdot \bb{\tilde{\varphi}}_\bb{k} \rangle_\psi - \langle\bb{G}\rangle_\psi = 0.
\end{equation}
Formally, this equation is over-determined: \eqref{eqn:solvability} provides $N_x$ equations for $N_x-1$ unknowns ($\tilde{\varphi}_{k_x\neq 0,k_y = 0}(z)$). While \eqref{eqn:solvability} can be made to be satisfied by the initial conditions, errors are introduced by the coupling of multiple flux-tube simulations needed for the novel radial boundary conditions; 
the  gyroaveraging of $\tilde{g}_{\bb{k},s}$ and $\tilde{\varphi}_\bb{k}$ that appears in~\eqref{eqn:quasi_rad} will sample points within the physical and boundary regions of the central domain when performed near the boundary. This results in discrepancies of $\varphi_\bb{k}$ within the boundary regions between the central and auxiliary flux-tube domains, which  then leads to errors in~\eqref{eqn:solvability}. To circumvent this issue, we include a correction term 
in the quasineutrality equation as
\begin{equation}\label{eqn:quasi_rad_matrix_tweak}
   \mathsfbi{Q}\bcdot \bb{\tilde{\varphi}}_\bb{k} -  \bb{G} - 
   \langle\langle\mathsfbi{Q}\bcdot \bb{\tilde{\varphi}}_\bb{k} -  \bb{G} \rangle_\psi\rangle_{\bb{x}_\perp} = -\mathsfbi{C}\bcdot ({\bb{\tilde{\varphi}}_\bb{k}}-\ba{\bb{\tilde{\varphi}}_\bb{k}}_\psi),
\end{equation}
where
\begin{equation}\label{eqn:perp_areal_average}
    \ba{A_\bb{k}}_{\bb{x}_\perp} = \fourier\left( \frac{\mr{TH}(\tilde{x})}{\ell_y(\ell_x - 2\tilde{L}_\mr{boundary})}\int_0^{\ell_y}\od \tilde{y} \int_{\tilde{x}_- + \tilde{L}_\mr{boundary}}^{\tilde{x}_+ - \tilde{L}_\mr{boundary}}\od \tilde{x}\,\fourier^{-1}\left(A_\bb{k}\right)\right)
\end{equation}
is the perpendicular areal average within the physical region, and
\begin{equation}
    \mr{TH}(\tilde{x}) = \left\{\begin{matrix}
     1 & \tilde{x}_- + \tilde{L}_\mr{boundary}< \tilde{x} < \tilde{x}_+ - \tilde{L}_\mr{boundary}, \\ 0 & \textrm{otherwise},
     \end{matrix}\right.
\end{equation}
is a top-hat function that excludes the boundary region. 
Equation~\eqref{eqn:quasi_rad_matrix_tweak} can then be solved after imposing a gauge potential, which we choose to be $\bb{\hat{k}}_0\bcdot \langle\mathsfbi{C}\bcdot\bb{\hat{k}}_0\bb{\hat{k}}_0\bcdot  \bb{\tilde{\varphi}}_\bb{k}\rangle_\psi = 0$, where $\bb{\hat{k}}_0$ is the unit column vector for the $\bb{k}=\bb{0}$ modes. The coupling of the $\bb{k} = \bb{0}$ modes to and from modes with $\bb{k}\ne \bb{0}$ is $\mathcal{O}(\Delta)$, and so the effect of the $\bb{k} = \bb{0}$ modes on physical observables is $\mathcal{O}(\Delta^2)$.  Numerical tests confirm the scaling of $|\tilde{\varphi}_{\bb{k} =\bb{0}}| \sim \Delta$, and so we expect the impact of the correction introduced in~\eqref{eqn:quasi_rad_matrix_tweak}  to be small.

\subsection{Sources and sinks}\label{sec:sources}

When the novel radial boundary condition is used for global simulations in \texttt{stella}, a mismatch of flux between the left and right simulation domains results in a pile-up or deficit of particles and heat in the central domain (the net flux of toroidal angular momentum in an up-down symmetric local simulation with no mean flow shear is zero, see~\citet{Parra_symmetry}). This necessitates the inclusion of a source or sink in the central domain. Global \texttt{stella} is equipped with two types of sinks: one based on a Krook-type operator and a new projection-based operator that exploits the scale-separated nature of our $\delta f$ approach.

The Krook type operator, commonly employed in global simulations, is of the form
\begin{equation}\label{eqn:krook_operator}
    D_\mr{K}(\tilde{g}_{\bb{k},s}) = - \nu_\mr{S} 
     \ba{\ba{\tilde{g}^\mr{even}_{\bb{k},s} - \frac{\tilde{F}_s (\rcoord)}{\tilde{n}_s(\rcoord)} \ba{\int \od^3 \tilde{v} \, J_{0s}\tilde{g}^\mr{even}_{\bb{k},s}}_\psi}_{\bb{x}_\perp}}_t, 
\end{equation}
where $\nu_\mr{S}$ sets the strength of the source operator, $\tilde{F}_s = (v^3_\mr{{th}s}/n_\mr{ref})F_s$, 
\begin{equation}
    \tilde{g}^\mr{even}_{\bb{k},s} = \frac{1}{2} \left[\tilde{g}_{\bb{k},s}(v_\parallel) + \tilde{g}_{\bb{k},s}(-v_\parallel)\right]
\end{equation}
is the even-in-$v_\parallel$ component of the distribution function, $\langle \cdots \rangle_{\bb{x}_\perp}$ is given by~\eqref{eqn:perp_areal_average}, and
\begin{equation}\label{eqn:time_average}
    \ba{\cdots }_t \doteq \frac{\int_0^t \od t' \exp(t'/\tau_\mr{S})(\cdots)}{\int_0^t \od t' \exp(t'/\tau_\mr{S})}
\end{equation}
is an exponentially-weighted time average, where $\tau_\mr{S}$ sets the averaging window~\citep{Candy_globloc}.  The Krook operator given by~\eqref{eqn:krook_operator} is similar to that advocated by~\citet{McMillan_sources}, where the even-in-$v_\parallel$ portion of $\tilde{g}_{\bb{k},s}$ is used in order to conserve momentum, and the zeroth moment of $\tilde{g}_{\bb{k},s}$ is subtracted in order to avoid introducing density perturbations on a flux surface. 
This operator is then included as an additional term on the right-hand-side of~\eqref{eqn:GKE_FT}, and operates only on the $k_y = 0$ modes. This, along with the time-averaging, ensures that this operator only affects the large-scale, long-time build up of particles and energy. The time-averaging procedure~\eqref{eqn:time_average} can be implemented numerically in a way that minimizes storage requirements, necessitating  only an additional array the size of $\tilde{g}_{\bb{k}_y =0,s}$ that conglomerates all the information needed from the previous timesteps.

The projection-operator-based sink,  provided as a more physically-motivated alternative to the standard Krook operator, is rooted in the $\delta f$ gyrokinetic formalism. Here, the full  distribution function $f$ is decomposed into a large-scale, long-time component $F$ and a small-scale, short-time component $\delta f$.  There then exists a suitable large-scale, long-time transport average $\ba{\cdots}_\mr{T}$ such that $\ba{f}_\mr{T} = F$ and $\ba{\delta f}_\mr{T} = 0$.  The $\delta f$ gyrokinetic equation can then be obtained from
\begin{equation}
    \frac{\partial \delta f}{\partial t} = \frac{\partial f}{\partial t} - \ba{\frac{\partial f}{\partial t}}_\mr{T}.
\end{equation}
It is this concept on which we base the projection-operator sink. Like the Krook-operator above, we let $\ba{\cdots}_\mr{T} =\langle \langle \cdots \rangle_{\bb{x}_\perp}\rangle_t$. This operator is then applied to the entire right-hand-side of the gyrokinetic equation, resulting in
\begin{align}\label{eqn:GKE_proj}
\frac{\partial g_s}{\partial t} = &- v_\parallel {\eb \bcdot \grad z} \left(\frac{\partial g_s}{\partial z}  + {\frac{Z_se}{T_s} F_s} \frac{\partial \langle \varphi \rangle_\bb{R}}{\partial z} \right)  + \frac{\mu_s}{m_s} {\eb \bcdot \grad B} \frac{\partial g_s}{\partial v_\parallel} \nonumber - {\bb{v}_{\mr{M}s}}\bcdot \left(\grad_\perp g_s +{\frac{Z_se}{T_s}F_s}\grad_\perp \langle \varphi\rangle_\bb{R}\right) \\
&- {\langle\bb{v}_\bb{E} \rangle_\bb{R} \bcdot \grad_\perp} g_s - {\langle \bb{v}_\bb{E} \rangle_\bb{R}\bcdot \grad\big|_E F_s}  - \Bigg\langle- v_\parallel {\eb \bcdot \grad z} \left(\frac{\partial g_s}{\partial z}  + {\frac{Z_se}{T_s} F_s} \frac{\partial \langle \varphi \rangle_\bb{R}}{\partial z} \right)  + \frac{\mu_s}{m_s} {\eb \bcdot \grad B} \frac{\partial g_s}{\partial v_\parallel}  \nonumber \\
&  - {\bb{v}_{\mr{M}s}}\bcdot \left(\grad_\perp g_s +{\frac{Z_se}{T_s}F_s}\grad_\perp \langle \varphi\rangle_\bb{R}\right) - {\langle\bb{v}_\bb{E} \rangle_\bb{R} \bcdot \grad_\perp} g_s - {\langle \bb{v}_\bb{E} \rangle_\bb{R}\bcdot \grad\big|_E F_s}   \Bigg\rangle_\mr{T}.
\end{align}
This operator is simple to implement numerically: one first calculates the distribution function for the next timestep  $\tilde{g}^{n+1}_{\bb{k},s, \mr{int}}$ as one normally would without any sinks, and then the actual distribution function at the next time step $\tilde{g}^{n+1}_{\bb{k},s}$ is given by
\begin{equation}
    \tilde{g}^{n+1}_{\bb{k},s} = \tilde{g}^{n+1}_{\bb{k},s, \mr{int}} - \upDelta t^n\ba{(\tilde{g}^{i+1}_{\bb{k},s, \mr{int}}-\tilde{g}^{i}_{\bb{k},s})/\upDelta t^i}_{\mr{T},\, i\,\in\, [0, n]},
\end{equation}
where $\upDelta t^i$ is the (possibly time dependent) simulation time step at step $i$ separating $\tilde{g}^i_{\bb{k},s}$ and $\tilde{g}^{i+1}_{\bb{k},s}$.
Apart from being a more physically motivated sink, this method also has the advantage of only requiring one parameter to be specified, $\tau_\mr{S}$, which must be chosen to be longer than any microscopic time-scale of interest; in practice, we find for global simulations that $\Delta^{-1}(a/v_\mr{thi}) \lesssim \tau_\mr{S} \lesssim \rho_\ast^{-1}(a/v_\mr{thi})$ is sufficient for convergence.

When employing the multiple-flux-tube radial boundary condition, discrepancies between the turbulence supplied by the auxiliary simulations and that of the central domain may sometimes result in strong shear layers just outside the boundary region. In order to mitigate these shear layers, global~\texttt{stella} supplies localized Krook-type sinks which can operate inside a subregion of length $L_\mr{K}$ of the boundary region in order to smooth the transition between the supplied turbulence and the modelled turbulence in the central region. These operators take the form
\begin{equation}
    D_\mr{BC}[\tilde{g}_{k_ys}(\tilde{x})] =  \left\{\begin{matrix} 
   \eta(\tilde{x}-\tilde{x}_- - \tilde{L}_\mr{boundary})(g_{s,\mr{L}}(\tilde{x}) - \tilde{g}_s(\tilde{x})) & \quad & \tilde{x}_- + (\tilde{L}_\mr{boundary} -L_\mr{K})< \tilde{x} < \tilde{x}_-  + \tilde{L}_\mr{boundary} \\
    \eta(\tilde{x}-\tilde{x}_++\tilde{L}_\mr{boundary})(g_{s,\mr{R}}(\tilde{x}) - \tilde{g}_s(\tilde{x})) & \quad & \tilde{x}_+ - \tilde{L}_\mr{boundary} < \tilde{x} < \tilde{x}_+  - (\tilde{L}_\mr{boundary}-L_\mr{K}) \\   
    0 & \quad & \textrm{elsewhere}\end{matrix} \right.
\end{equation}
where $\eta(x)  = \nu_\mr{BC}[ 1-(1-\rme^{-3(L_\mr{K} -|x|)/L_\mr{K}})/(1-\rme^{-3})]$ and $\nu_\mr{BC}$ sets the strength of the operator. In practice, $\nu_\mr{BC}$ is chosen to be comparable to the growth rate of the fastest growing mode over all three domains. When these Krook operators are used in simulations with the novel boundary condition, $g_\mr{s,\mr{L}}$ and $g_\mr{s,\mr{R}}$ are only directly copied into the Krook-free portion of the boundary region. Like the case with $\tilde{L}_\mr{boundary}$, the size of the Krook damping region should be similar to the size of the largest Larmor orbit contained in the simulation.

\subsection{Equilibrium flow shear}\label{sec:flowshear}

The local version of \texttt{stella} currently implements equilibrium flow shear using the discrete wavenumber-shift method~ formulated by \citet{Hammett_FlowShear} and includes the nonlinear corrections advocated by~\citet{McMillan_FlowShear} which renders the scheme semi-continuous.

The wavenumber advection method is based on the fact that for a linear background velocity profile $v_{\bb{E},0,y} = \gamma_E\tilde{x}$, where $\gamma_E$ is the shear rate, the advection equation
\begin{equation}
\frac{\partial g}{\partial t} + v_{\bb{E},0,y}(x) \frac{\partial g}{\partial y} + \ldots
\end{equation}
 is equivalent to solving the system with a time-dependent $k_x$,
\begin{equation}\label{flowshear}
\frac{\partial }{\partial t}g(k_x - k_y \gamma_E t,k_y) + \ldots.
\end{equation}
For gyrokinetic codes employing an implicit integration scheme, having a time-dependent $k_x$ would require a recalculation of the response matrices at every simulation time step, and thus would be prohibitively expensive computationally. The wavenumber shift approach instead remaps the distribution function discretely.
To determine when and how this remapping occurs, we define a $k_y$-dependent marker wavenumber $k_x^\ast(t) \doteq - k_y \gamma_E t$ that evolves in time. Whenever $|k_x^\ast(t) - k_x^\ast(t^*_\mr{prev})| > \upDelta k_x /2$, where
\begin{equation}
    t^*_\mr{prev} = \upDelta t_E \left[\operatorname{floor}\left(\frac{t_\mr{prev}+\upDelta t_E/2}{\upDelta t_E}\right) - \frac{1}{2}\right], 
\end{equation}
 $t_\mr{prev}(k_y)$ is the time of the previous remapping for binormal wavenumber $k_y$, $\upDelta t_E = |\upDelta k_x/ \gamma_E k_y|$ is the nominal time between remappings for binormal wavenumber $k_y$, and $\upDelta k_x$ is the radial wavenumber grid spacing, then  information of the distribution function $g_{\bb{k},s}$ at radial wavenumber $k_x$ is transferred to radial wavenumber $k_x - \operatorname{sign}(\gamma_E)n_\mr{shift}\upDelta k_x$, i.e.,
\begin{equation}
    g^{n+1}_{k_x - \operatorname{sign}(\gamma_E)n_\mr{shift}\upDelta k_x, k_y} = g^{n}_{k_x, k_y},
\end{equation}
where $n_\mr{shift} = \mr{round}[|k_x^\ast(t) - k_x^\ast(t^*_\mr{prev})|/\upDelta k_x]$ is the number of radial wavenumbers by which the distribution function is to be shifted. This remapping, if it is to occur, is done at the beginning of a simulation time step before any other term in the gyrokinetic equation is calculated. This scheme, which leaves the radial wavenumber grid unchanged, ensures that $k_x$ on the fixed grid is as close to the continuous time $k_x(t) = k_x(t=0) + k_x^\ast (t)$ as possible. 
Note that the $k_y=0$ zonal modes are unaffected by this shearing, and so do not require remapping. 

It has been recently shown that the accuracy of the wavenumber advection scheme can be improved by incorporating a continuously varying $k_x$, as well as a continuous flow shear, into the numerical integration of the nonlinear terms that appear in the gyrokinetic equation~\citep{McMillan_FlowShear,christen_2021}. These modifications are done to avoid the problem where modes at different $k_y$ that are initially uncoupled nonlinearly become coupled due to staggered remapping. As the nonlinear terms are typically integrated using an explicit scheme, these modifications can be made at little extra computational cost, and are implemented in a straightforward manner: first,  the substitution $k_x \rightarrow k_x - k_y\gamma_E(t - t_\mr{prev})$ is made for every $k_x$ appearing in the nonlinear term.\footnote{This substitution is not made in the $J_0$ Bessel functions that appear in the nonlinear term. This is not needed, however, to avoid spurious cross-coupling between uncorrelated modes.} Second, a phase-shift to $g_{\bb{k},s}$ and $\varphi_\bb{k}$ is applied by multiplying an appropriate prefactor, 
\begin{equation}\label{eqn:prefactor}
    \varphi_{k_y}(x) \rightarrow \varphi_{k_y}(x) \exp[-\imag k_y \gamma_E x(t-t^*_\mr{prev})].
\end{equation}
This renders the effects of shearing semi-continuous. Once the nonlinear term is calculated, the result is divided by this prefactor, and the numerical integration then proceeds as usual.

The global version of \texttt{stella} employs a continuous form of equilibrium flow shear, which simply adds flow shear as a background linear profile.  While this approach has been used in gyrofluid simulation in the past~\citep{Waltz_1994}, it has the disadvantage of introducing a region of very strong shear at the radial boundaries of the simulation where the shear profile experiences a discontinuity, and thus results in excessive levels of turbulence stabilization. This drawback can be avoided entirely by utilizing the novel boundary conditions detailed in \S \ref{sec:radBC}; as the boundary regions are replenished with `good' information at every time step, the effects of the large amounts of shear at the profile discontinuities is reduced. When this approach is used, the shear rate is made continuous across all three domains by using the wavenumber shift approach, along with nonlinear corrections, in the left and right domains.  Prior to communicating information of the left and right distribution functions, $g_{s,\mr{L}}$ and $g_{s,\mr{R}}$ are multiplied by the phase factor given in~\eqref{eqn:prefactor} in order to render the shearing semi-continuous. These distribution functions are then Doppler shifted in order to match the velocities at the boundaries of the central domain. After these steps are performed, the information can be copied as usual.

{
Like the spatial operator $\spaceOperator$ detailed in \S \ref{sec:sim_equations}, the continuous approach to flow-shear space suffers the drawback of wrap-around, where a mode at one extreme of the radial wavenumber grid gets transferred to the opposite extreme, rather than moving off-grid. 
In the discrete wavenumber shift approach, which is done in $k_x$-$k_y$ space,  the wrap-around problem is easily avoided: information of the distribution function that gets sheared off the radial wavenumber grid is simply zeroed out. In the global version of equilibrium shear, which is done in $x$-$k_y$ space, a mode at $k_{x,\mr{max}}$ would unphysically wrap around to $-k_{x,\mr{max}}$ if no additional steps were taken. Unlike the shearing resulting from the $\rho_\ast$-small terms due to radial profile variation, the shearing from the equilibrium flow is of order unity, and so one cannot rely on hyper-dissipation alone to prevent wrap-around. Fortunately, the advection speed $v_{k_x} = - \gamma_E k_y$ due to equilibrium flow shear is well defined for every $k_y$,  and is uniform across the entire $z$ domain.  We thus prevent wrap-around with global shear by zeroing out the distribution function at $k_x = \pm k_{x,\mr{max}}$ over the same $k_y$-dependent frequency that a sheared mode takes to cross a single cell in the radial wavenumber grid, i.e., whenever $|k_x^\ast(t) - k_x^\ast(t^{**}_\mr{prev})| > \upDelta k_x$, where $t^{**}_\mr{prev} = \upDelta t_E \operatorname{floor}(t_\mr{prev}/\upDelta t_E)$ and $t_\mr{prev}$ here refers to the last time a mode at $k_y$ was zeroed out. We have found that this largely eliminates any wrap-around of information across the boundaries of the $k_x$ grid.
}

\subsection{Hyper dissipation}\label{sec:hyperdissipation}

A fourth-order hyper-dissipation operator is used in \texttt{stella} in order to prevent a pile-up of energy at small perpendicular scales due to turbulent cascade. This operator, which is applied to the distribution function as an additional implicit step in the operator splitting scheme, has the form 
\begin{equation}
 g_{n+1} = \frac{1}{1+ \upDelta t D_\mr{hyper}[k^2_{\perp,\textrm{hyper}}/\operatorname{max}(k^2_{\perp,\textrm{hyper}})]^2}g_n,
\end{equation}
where $D_\mr{hyper}$ is an adjustable coefficient which sets the strength of the dissipation and $k^2_{\perp,\rm{hyper}}(k_x,k_y)$ is some measure of the squared perpendicular wavenumber. For local \texttt{stella}, 
\begin{equation}
    k^2_{\perp,\mr{hyper}} = k_\perp^2 = k_x^2 |\grad x|^2 + 2  k_xk_y(\grad x\bcdot \grad y) + k^2_y |\grad y|^2 
\end{equation}
is the physical squared perpendicular wavenumber. The geometrical coefficients $|\grad x|^2$, $\grad x\bcdot \grad y$, and $|\grad y|^2$ vary in $\vartheta$, and ensure continuity across the parallel boundary for modes connected by the parallel boundary condition (\S  \ref{sec:coords}). These coefficients, however, also vary radially along with the geometric profile variation, thus complicating the application of hyper-dissipation for global simulations. This difficulty can be avoided by using an alternative definition for $k_{\perp,\mr{hyper}}$:
\begin{equation}
    k^2_{\perp,\mr{hyper}} =  \left\{ \begin{matrix}k_x^2 & \quad & k_y = 0, \\ k_y^2[1+\lambda(\theta - \theta_0)^2]& \quad & k_y \ne 0,\end{matrix}\right.
\end{equation}
where $\theta_0 \doteq k_x/k_y$ is the ballooning angle and $\lambda = 1$ when $q$ is used as the $x$ coordinate.\footnote{Our definition of the ballooning angle differs from the standard one, $\theta_0 = k_x/k_y \hat{s}$, as we use $q$ as the radial coordinate instead of $\psi$.} This form of $  k^2_{\perp,\mr{hyper}}$ retains the $\theta$ dependence that ensures continuity across the parallel boundary, while doing away with the radial variation of the geometrical coefficients. In practice, the results of a nonlinear simulation should be insensitive to the exact form of the hyper-dissipation operator, provided the small perpendicular scales are well resolved.


\section{Numerical benchmarks}\label{sec:tests}

In this section we perform a variety of tests in order to demonstrate the efficacy of our new approach to global gyrokinetics; namely, that of the novel radial boundary conditions and the inclusion of next-order corrections in the underlying gyrokinetic equations. Unless otherwise noted, all simulations  used standard Cyclone Base Case (CBC) parameters which are tabulated in Table~\ref{tab:CBC_param}, and we define $\rho_\ast = \rho_\mr{ref}/a$. Additionally, the  values $N_x$ and $N_y$ refer to the number of modes resolved in the radial and binormal direction after $2/3$ dealiasing, and so the number of collocation points when calculating the nonlinear terms is effectively larger by a factor of $3/2$.  This dealiasing is not performed during the communication of the novel radial boundary condition, and so the boundary size $N_\mr{boundary}$ should be compared with $N_x$, rather than $3N_x/2$.

\begin{table}
    \centering
        \begin{tabular}{c|c}
    input parameter & value \\
    \hline
        $m_\mr{i}/m_\mr{ref}$ & 1 \\     
        $n_\mr{i}(\rcoord_0)/n_\mr{ref}$ & 1 \\        $T_\mr{i}(\rcoord_0)/T_\mr{ref}$ & 1 \\
        $n_\mr{i}/ n_\mr{e}$ & 1 \\
        $T_\mr{i}/T_\mr{e}$ & 1 \\
        $R_0/L_{n_\mr{i}} = -(R_0/n_\mr{i}) n'_\mr{i}$ & 2.2 \\
        $R_0/L_{T_\mr{i}} = -(R_0/T_\mr{i}) T'_\mr{i}$ & 6.9 \\
         $\epsilon = \rcoord_0/a $ & 0.5\\
        $R_0/a$ & 2.77778 \\
    \end{tabular}
    \hspace{1cm}
    \begin{tabular}{c|c}
    input parameter & value \\
    \hline
       $q$  & 1.4 \\
        $\hat{s}$ & 0.796 \\
        $\kappa$ & 1 \\
        $\kappa'$ & 0 \\
        $\delta$ & 0 \\
        $\delta'$ & 0 \\
        $\beta$ & 0 \\
        $\beta'$ & 0 \\
        \\
    \end{tabular}
    \caption{Parameters of the Cyclone Base Case (CBC).}
    \label{tab:CBC_param}
\end{table}

\subsection{Continuity of the radial boundary conditions}

\begin{figure}
    \centering
    \includegraphics[width=\textwidth]{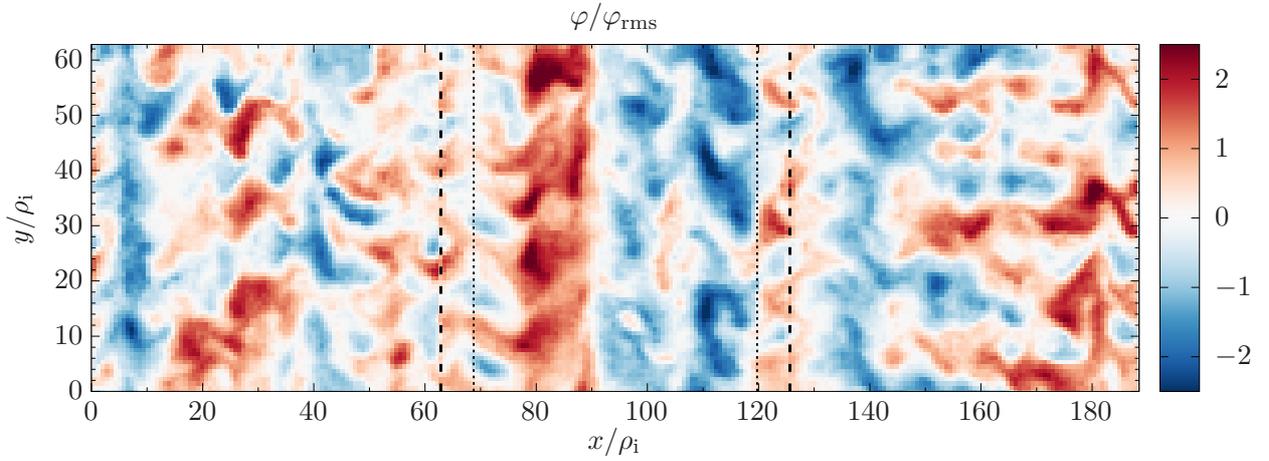}
    \caption{Snapshot of the electrostatic potential in the perpendicular plane situated at the outboard midplane for a simulation using CBC parameters with no radial profile variation.  Dashed lines denote the simulation domain boundaries.}
    \label{fig:CBC_omp_section}
\end{figure}

\begin{figure}
    \centering
    \includegraphics{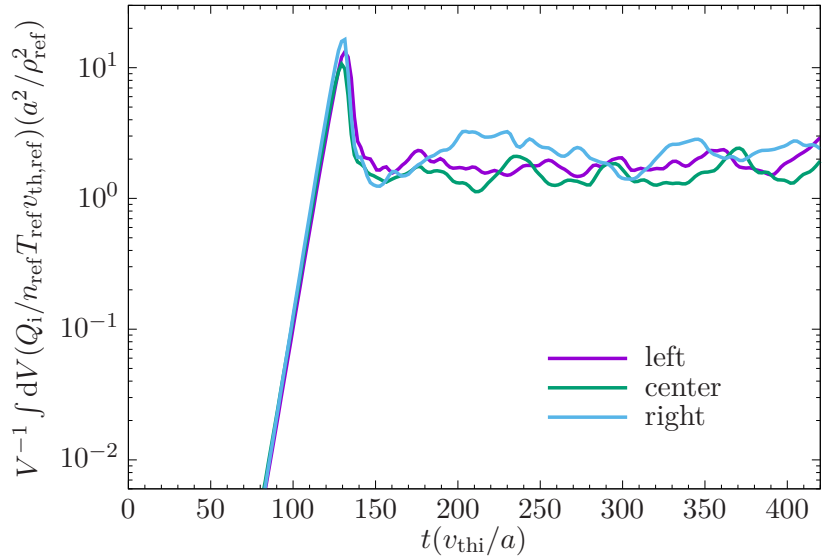}
    \caption{Ion heat fluxes in each domain of a multiple-domain nonlinear simulation without radial profile variation ($\rho_\ast =0$).}
    \label{fig:CBC_fluxes}
\end{figure}

To test the performance of the new radial boundary conditions, we perform a simulation using CBC parameters without any profile variation  ($\rho_\ast = \Delta = 0$) in order to demonstrate continuity between the physical domains. This simulation uses adiabatic electrons, 85 dealiased modes in each of the the radial and binormal directions, 32 points along the magnetic field, 24 points in the parallel velocity space, 6 points in $\mu$ space, and $D_\mr{hyper} = 0.05$. The boundary regions in the center domain each use ten co-location points, and the boundary Krook operator is applied in the inner eight points with $\nu_\mr{K}(a/v_\mr{th,ref}) = 0.5$. Each flux-tube domain is seeded with small amplitude random noise, and after a period of exponential growth in the initial linear regime, the nonlinearity becomes dynamically important and the domains reach a saturated turbulent state.  No other sources or sinks are used in these simulations. 

Figure~\ref{fig:CBC_omp_section} shows a snapshot of the electrostatic potential for each simulation domain in the perpendicular plane situated at the outboard midplane. As no radial profile variation is included in these simulations, the statistical characteristics of the turbulence in each simulation domain should be identical to one-another. Indeed, the cross-section of $\varphi$ shows similar turbulence across all the domains, without any spurious behaviour occurring near the domain boundaries. The heat fluxes in each domain are also shown in figure~\ref{fig:CBC_fluxes} to be statistically identical, and the linear growth rates (not shown) are also consistent. The novel radial boundary condition is thus shown to be well behaved. An expression for the heat flux is given in~\ref{app:fluxes}.

\subsection{Equilibrium flow shear}

Our global implementation of equilibrium flow shear is tested by comparing the resulting fluxes to those obtained by local flux-tube simulations using the standard wavenumber shift method~\citep{Hammett_FlowShear} along with non-linear corrections~\citep{McMillan_FlowShear}.  Unlike radial profile variation, equilibrium flow shear is not a $\rho_\ast$-small effect, and so these tests are performed without other global effects (i.e. $\rho_\ast = \Delta = 0$).

In these tests, our global simulations use the CBC parameters and a resolution of $(N_y,N_z,N_{v_\parallel}, N_\mu) = (21,16,24,6)$. The side domains use $N_x = 43$ with a radial extent of $\ell_x = 62.8\rho_\mr{i}$ while the central domain uses $N_x = 85$ dealiased modes with a radial extent of $\ell_x = 125 \rho_\mr{i}$. The boundary regions use 16 cells each, with a Krook operator applied on the outer 10 cells with a damping rate of $0.2$. If the Krook operator is not used, then shear layers can develop at the interfaces of the boundary region which may counteract the effect of the background flow shear. The local simulations use the same parameters as the central domain in the global simulation, and adiabatic electrons are employed throughout.  No other sources or sinks are used in the simulations.

\begin{figure}
    \centering
    \includegraphics[width=\textwidth]{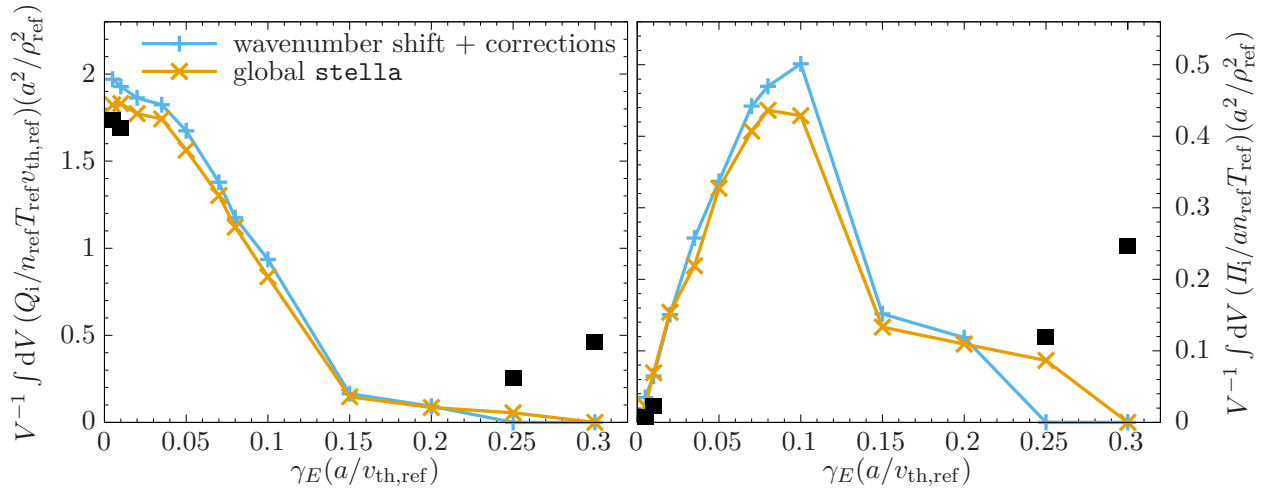}
    \caption{Comparison of the total heat flux (left) and momentum flux (right) between a local flux-tube simulation (yellow line) and a global simulation (blue line) with no profile variation ($\rho_\ast = 0$) across a range of shear rates $\gamma_E$.  Each simulation employs both parallel and perpendicular flow shear. Black squares denote results from the global simulation without the zeroing out described in the last paragraph of \S \ref{sec:flowshear}.}
    \label{fig:shear_both_LR}
\end{figure}

Figure~\ref{fig:shear_both_LR} shows the total heat flux (left) and parallel momentum flux (right) for the global implementation of flow shear (yellow line) along with the result from local flux-tube simulations (blue line). Expressions for these fluxes are given in \ref{app:fluxes}. These results reveal that our global implementation of flow shear compares well to the more standard method.  { Also included in this plot as black squares are the results from simulations of using global flow shear without the periodical zeroing-out of information at the boundary. As argued in \S \ref{sec:sim_equations} and \S \ref{sec:flowshear}, the results are insensitive to this zeroing for small values of shearing due to the presence of finite hyper-dissipation. However, for large flow shear the information can escape this hyper-dissipation, thus causing an over-prediction of both heat and momentum flux  when the zeroing-out dealiasing routine is not performed; this over-prediction worsens as shearing is increased.}

\subsection{Consistency of the first derivatives}

\begin{figure}
    \centering
    \includegraphics{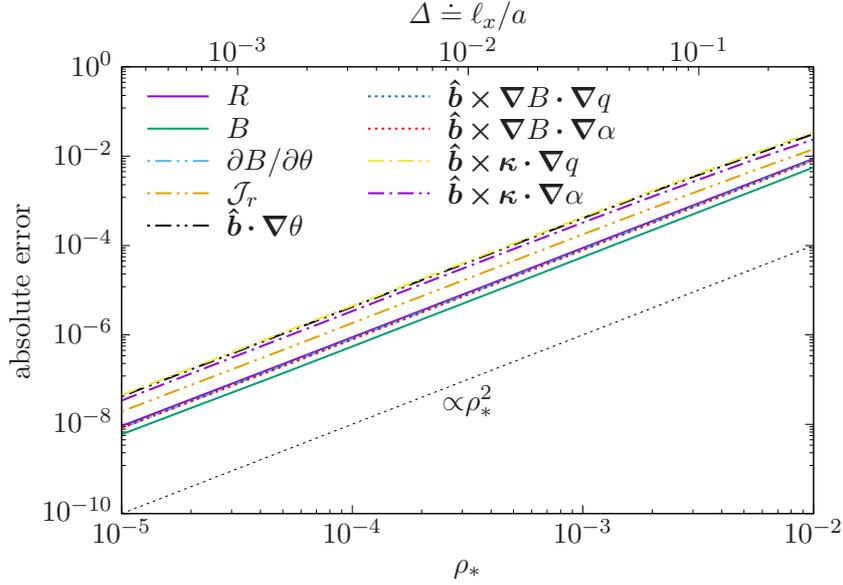}
    \caption{Convergence of several geometric factors as a function of $\rho_\ast$. Solid lines denote the line-averaged absolute error, e.g. $\delta R = L_z^{-1}\int \od z |R(r + \upDelta r, z) - R(r,z) - \upDelta r R'(r,z)|$. Dotted line denotes the ideal scaling $\rho_\ast^2$. Here, $\rho_\ast = \rho_\mr{ref} / a$.
    }
    \label{fig:milerr}
\end{figure}

Our approach to global gyrokinetics involves the Taylor expansion of the geometrical coefficients (given in~\ref{app:mag_geo}), as well as the density and temperature profiles. This introduces various additional terms in the gyrokinetic and quasineutrality equations, as well as the gyro-averaging operator. In order to verify the implementation of these terms, we compare the geometrical coefficients computed in two separate local flux-tube simulations; one simulation is performed at a displaced radial location $\rcoord = \rcoord_0 +  \upDelta \rcoord_+$ that does not employ the additional terms in the Taylor expansion, and a simulation located at $\rcoord = \rcoord_0$ which then employs these additional terms to evaluate the coefficients at $\rcoord_0 + \upDelta \rcoord_+$ (thus placing the flux tube at $\rcoord=\rcoord_0 + \upDelta r_+$). We then compute the absolute error between the geometrical coefficients of the two simulations and plot this as a function of $\rho_\ast$ (equivalently, $\Delta$); proper implementation of the terms resulting from the first-order Taylor expansion results in a scaling of the absolute error with $\rho_\ast^2$. Plotted in figure~\ref{fig:milerr} is the error scaling of various geometrical coefficients with $\rho_\ast$ which do indeed exhibit a $\rho_\ast^2$ scaling for a broad range of $\rho_\ast$. While this figure only displays a few geometrical coefficients, these scalings have also been verified for all other geometrical coefficients, as well as for terms resulting from radial variation in density and pressure. As a final note, the error scalings for some quantities in figure~\ref{fig:milerr} do break from the $\rho_\ast^2$ scaling for $\Delta \gtrsim 0.3$; however, for these order-unity values of $\Delta$ the Taylor expansion is expected to breakdown.

\subsection{Convergence to the $(k_\perp L)^{-1} \rightarrow 0$ limit}

\begin{figure}
    \centering
    \includegraphics[width=\textwidth]{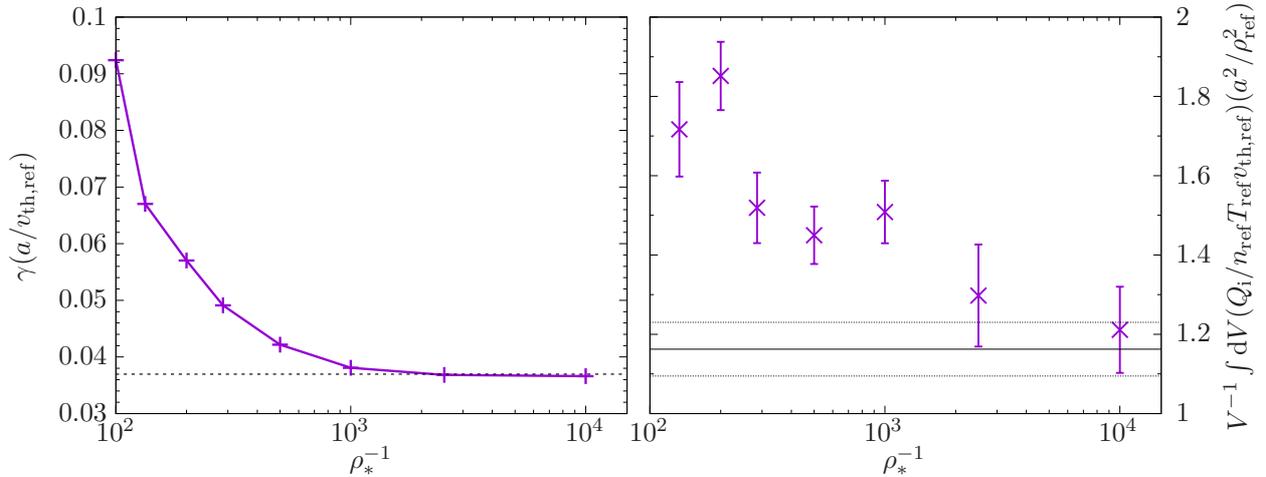}
    \caption{(\emph{left}) Time-averaged growth rate of the volume-average ion heat flux as a function of $\rho_\ast$ for radially global linear simulations. Dotted line indicates the growth rate of a local linear simulation centered at $r/a = 0.5$. (\emph{right}) Time- and volume-averaged ion heat flux in a saturated turbulent state for radially global nonlinear simulations. Error bars denote the standard error of the mean. Solid line indicates the time- and volume-averaged heat flux of a local nonlinear simulation centered at $r/a = 0.5$, while the dotted lines denote the standard error of the mean for that simulation. Here, $\rho_\ast = \rho_\mr{ref} / a$.}
    \label{fig:rhostar_scan}
\end{figure}

Local flux-tube gyrokinetics is obtained by formally taking the $(k_\perp L)^{-1} \rightarrow 0$ limit of the gyrokinetic equations~\citep{Parra_globallocal}. Thus, a stringent test that all global gyrokinetic codes must pass is to prove convergence to local gyrokinetic simulations in the $(k_\perp L)^{-1} \rightarrow 0$ limit, which can be done by letting $\Delta$ tend towards zero. We test this convergence in $\texttt{stella}$ using the Cyclone Base Case parameters with adiabatic electrons, as well as the additional second derivatives $ q''a^2 = 5$, $ n''(a^2/n) =-1 $, $ T''_\mr{i}(a^2/T_\mr{i}) =-4$. These second derivatives have been chosen in order to provide appreciable modification to the gradient scale lengths as $\Delta$ approaches order unity.
The multiple flux-tube boundary conditions are employed, and each simulation domain has a resolution of $(N_x, N_z,N_\mu, N_{v_\parallel}) = (43,24,12,48)$, a single poloidal turn and a radial extent in terms of gyroradii of $\ell_x = 60 \rho_\mr{i}$. The size of the radial boundary is 4 collocation points. The spatial resolution is held fixed between the simulations, and so as $\rho_\ast$ is increased, the physical region of the device that is sampled (as a portion of the minor radius $a$) is also increased. Sources or sinks are not needed in linear simulations since profile relaxation is a nonlinear effect, and so sources or sinks are not used in the linear simulations reported in this section. Each domain is seeded with a single $k_y\rho_\mr{i} = 0.2$ mode which, in a linear simulation, decouples from other modes with different binormal wavenumber. As the system evolves, the growth rate of the volume-averaged heat flux (\ref{app:fluxes}) eventually reaches a constant value, which is plotted in the left panel of figure~\ref{fig:rhostar_scan} as a function of $\rho_\ast$. The global simulations are indeed shown to converge to the local result, which is denoted in the figure with a dotted line. Apart from the largest value of $\rho_\ast$ at which the $\Delta$ expansion begins to breakdown, this growth rate is nearly a monotonically increasing function of $\rho_\ast$. This is due to the radial domain sampling regions of larger $R/L_T$ near the outer edge of the torus, an effect that disappears as $\rho_\ast$, and thus the radial extent in terms of $r$, is decreased. Additional simulations for $\rho_\ast = 0.002$ where $N_z$, $N_\mu$, and $N_{v_\parallel}$ are separately halved have been performed to ensure convergence of the growth rates.

We also perform a $(k_\perp L)^{-1} \rightarrow 0$ convergence test for nonlinear simulations by comparing the time- and volume-averaged ion heat flux of the saturated turbulent state. This is performed using the same parameters as in the linear simulations of the previous paragraph, but with the resolution of $(N_x, N_y, N_z,N_\mu, N_{v_\parallel}) = (171,85,16,8,48)$. Additionally, the simulations employ the projection-based source with $\tau_\mr{S}(v_\mr{th,ref}/a) = 50$,  and no boundary-region Krook operator.
The results of these simulations are shown on the right panel of figure~\ref{fig:rhostar_scan}, along with the result of a nonlinear flux-tube simulation performed at $\rcoord/a = 0.5$. While the heat flux signals are much noisier than the growth rates of the linear simulations, convergence in $\rho_\ast$ is indeed demonstrated for small enough $\rho_\ast$.

Finally, in order to showcase the utility of the hybrid global-local approach, we perform a $(k_\perp L)^{-1} \rightarrow 0$ convergence test where the physical region of the device that is sampled is held fixed, rather than the box size in terms of the number of gyroradii.  Three sets of simulations are used, employing the multiple-flux-tube boundary condition, the periodic-triangle-wave boundary condition detailed in~\citet{Candy_globloc}, as well are Dirichlet boundary condition. The numerical resolution for these simulations are $(N_y, N_z, N_\mu, N_{v_\parallel}) = (43,16,8,36)$. The radial resolution for the simulations employing the multiple-flux-tube and Dirichlet boundary conditions are $N_x = 43$, $171$, $341$ for $\rho_\ast^{-1} = 250$, $500$, $1000$, $2000$, respectively, while using a boundary size of four collocation points for $\rho_\ast^{-1} = 250$ and eight for the remaining three values of $\rho_\ast$. Additionally, the boundary Krook operator is applied to half of those points for these simulations. Dirichlet boundaries are applied similarly to the multiple-flux-tube approach, but with zeroes instead of information sourced from auxiliary local simulations. The simulations employing the periodic triangle wave boundary condition have radial box sizes and resolutions that are twice as large as the other two analogous cases, but do not need to use auxiliary flux-tube simulations or a boundary-region Krook operator. All simulations employ the Krook-based source with $\tau_\mr{S}(v_\mr{th,ref}/a) = 50$ and $\nu_\mr{K}(a/v_\mr{th,ref}) = 0.15$. We note that the periodic triangle simulations require twice the computational resources  to perform compared to analogous simulations employing Dirichlet boundary conditions, while the multiple flux-tube simulations requires three times as much, though this can be mitigated in the future by using smaller radial domains in the auxiliary simulations and appropriate load balancing.

\begin{figure}
    \centering
    \includegraphics[width=0.6\textwidth]{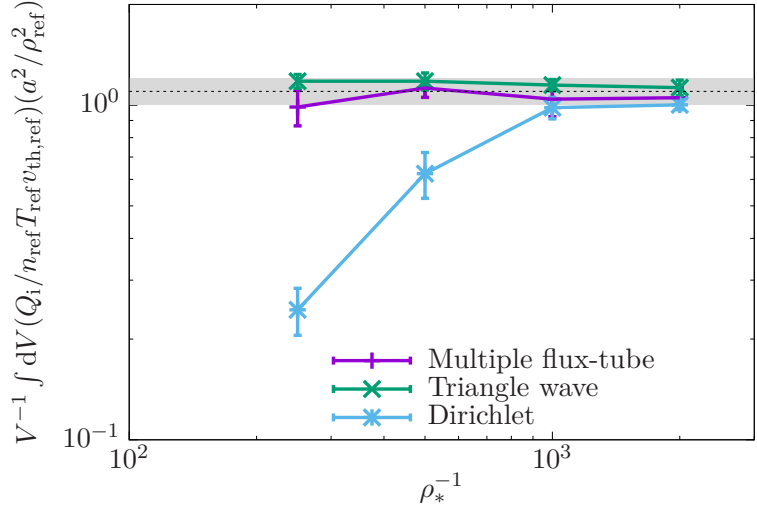}
    \caption{Ion heat fluxes as a function of $\rho_\ast$ for simulations where the sampled physical region is held fixed. Included are simulations which use a multiple-flux-tube radial boundary condition (purple line, \S \ref{sec:radBC}), the periodic triangle wave boundary condition (green line, \citep{Candy_globloc}), as well as Dirichlet boundary conditions (blue line). Fluxes are time averaged over the saturated states and volume averaged within the center half of the simulation domain. The dotted line is the result of a local simulation located at $\rcoord/a = 0.5$, while the shaded region denotes the standard error of that mean. }
    \label{fig:rhostar_scan_fixbox}
\end{figure}

The results from these simulations are given in figure~\ref{fig:rhostar_scan_fixbox}, which displays the ion heat flux as a function of $\rho_\ast$. Here, heat fluxes are time-averaged over the saturated state, while volume-averaged over the half-domain centered around $\rcoord/a = 0.5$; for the simulations employing periodic triangle wave boundary conditions, this is the half-domain of the physical region of the simulation, which is the left half of the simulation domain. We see from figure~\ref{fig:rhostar_scan_fixbox} that 
the simulations using multiple flux-tube and triangle wave boundary conditions agree with local result for $\rho_\ast$ at least as large as $\rho_\ast  \approx 1/250$, while the simulations employing Dirichlet boundary conditions significantly underpredict the fluxes for $\rho_\ast^{-1} \lesssim 500$.  This archetypal under-prediction of the fluxes relative to the local result is often attributed to global effects, such as profile shearing, that are absent in local codes: The results shown here instead indicate that the artificial damping of fluctuations associated with Dirichlet boundary conditions is responsible for the reduction in flux with $\rho_\ast$.  Such a concern has already been raised in~\citet{Candy_globloc}, who found a similar insensitivity of the heat flux to variation of $\rho_\ast$ when using the periodic-triangle-wave boundary condition.

\subsection{Rosenbluth-Hinton tests}

A standard benchmark for gyrokinetic codes is the Rosenbluth-Hinton test~\citep{Rosenbluth}, which verifies the implementation of the parallel streaming terms, the radial magnetic drifts and the $k_y = 0$ mode in the $k_\perp \rho_\mr{i}\rightarrow 0$ limit. In this test, the residual level of a zonal electrostatic perturbation that has been damped by collisionless processes is measured and compared to the theoretical prediction for large aspect ratio circular geometry:
\begin{equation}\label{eqn:RH_res}
   \mr{RH}_\mr{XC} \doteq \frac{\varphi(t)}{\varphi(0)} = \left[1 + q^2(1.64 + 0.5\sqrt{\epsilon} + 0.361 \epsilon)/\sqrt{\epsilon}\right]^{-1}.
\end{equation}
This residual level, originally calculated by~\citet{Rosenbluth}, was given to higher accuracy in $\epsilon$ and $q$ by~\citet{Xiao2006}. 

\begin{figure}
    \centering
    \includegraphics[width=\textwidth]{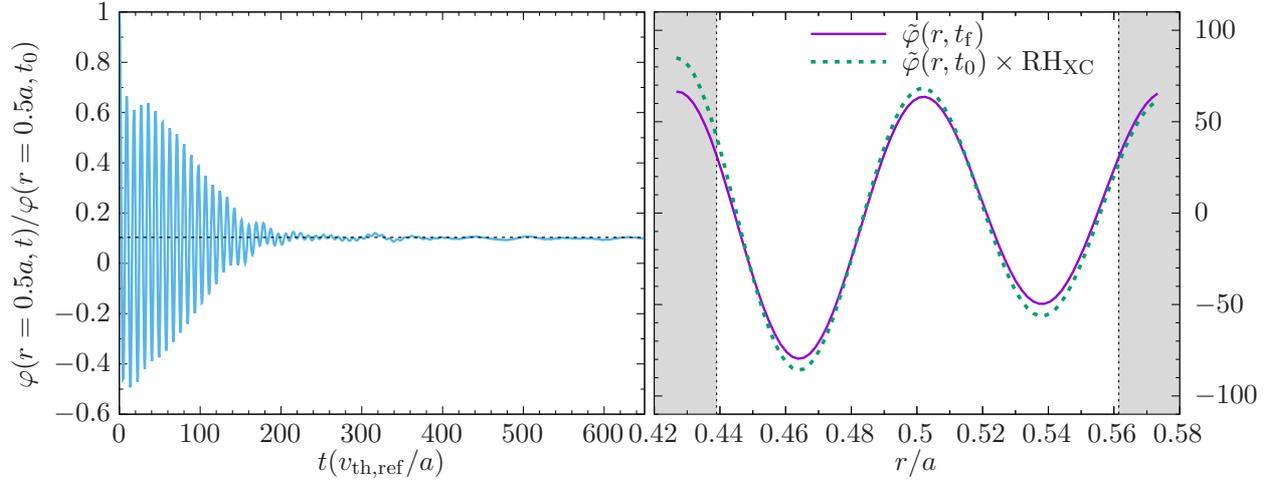}
    \caption{(\emph{left}) Rosenbluth-Hinton residual of the electrostatic potential $\varphi$ (solid line) at the center of the radial domain for a global simulation. Dotted line denotes expected residual level. (\emph{right}) Comparison of the (time-averaged) electrostatic potential in its final state (solid line) to the initial electrostatic potential scaled by the expected radially dependent  residual fraction (dotted line) as given by~\citet{Xiao2006}. Shaded area denotes the boundary region of the central flux tube.}
    \label{fig:RH_test}
\end{figure}

To perform the Rosenbluth-Hinton test using global \texttt{stella}, we run a multiple flux-tube linear simulation with velocity space resolution $(N_{v_\parallel}, N_\mu) = (192,6)$, $N_z = 128$, $N_x = 43$ for the side domains and $N_x = 85$ for the central domain. The radial extent of the side domains is $\ell_x= 300\rho_\mr{i}$ while for the center it is $\ell_x = 600\rho_\mr{i}$. 
Flat density and temperature profiles are used for all simulations, while the magnetic geometry is taken from the CBC, along with $q''a^2 = 0.5$ and $\rho_\ast = 0.0002$ ($\Delta = 0.12$). Each boundary region uses eight collocation points and no dissipation is used, and adiabatic electrons are employed. Simulations are initialized with a single large-wavelength $k_x\rho_\mr{i} = 2\uppi / 300$ mode, and after some time a residual level is eventually maintained. The left panel of figure~\ref{fig:RH_test} shows the residual level of the zonal flow at the center of the middle domain (at $\rho = 0.5$) along with the analytical prediction given by~\eqref{eqn:RH_res} (this denoted with a dotted line). It is clear that global \texttt{stella} captures the correct residual level at the center of the domain.

While the original Rosenbluth-Hinton calculation is performed in the local limit, we can let~\eqref{fig:RH_test} be a function of $r$ through $\epsilon$ and $q$ and ascertain how well \texttt{stella} attains this `global' version of the residual level.  The right panel of figure~\ref{fig:RH_test}  displays the final zonal potential at the end of the simulation (purple line), as well as the initial potential profile scaled by the predicted residual fraction given by~\eqref{eqn:RH_res} (green line). At least at this modest $\rho_\ast$, the global simulation captures the correct residual level over the entire domain. 
 
\subsection{Sources and sinks}

\begin{figure}
    \centering
    \includegraphics{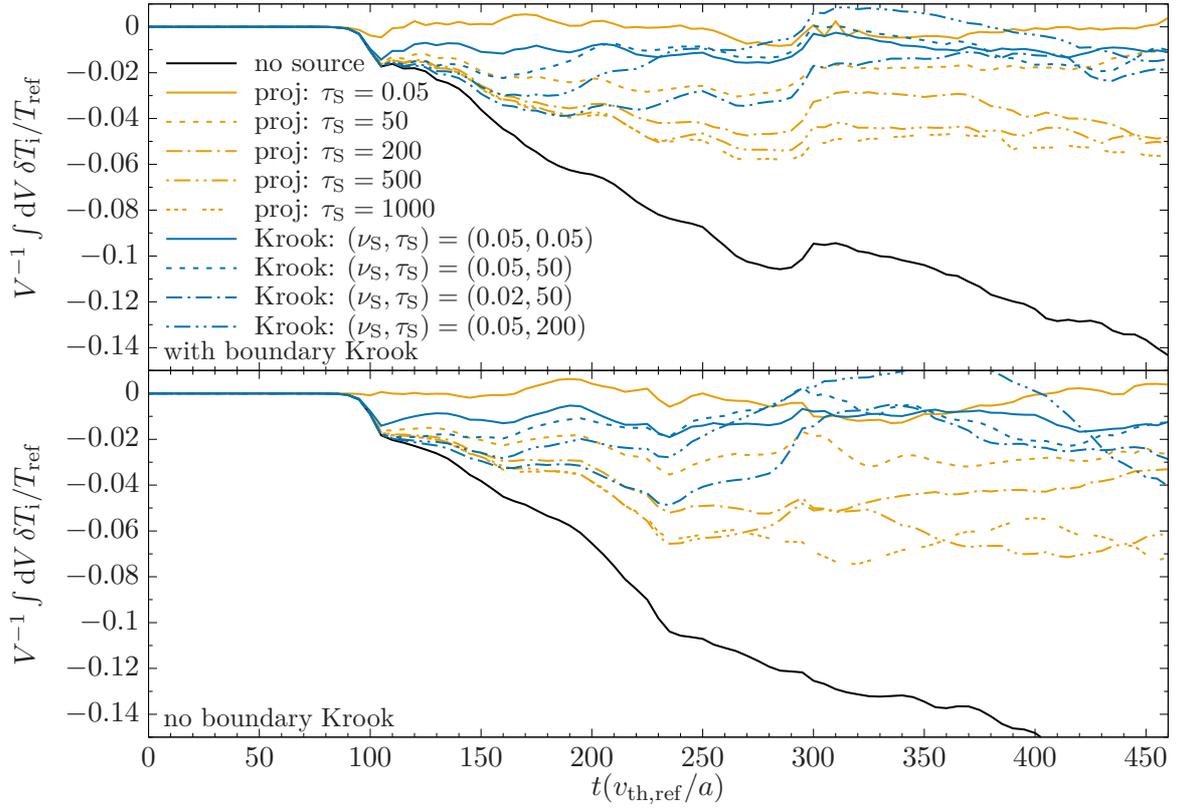}
    \caption{Evolution of the box-averaged temperature as a function of time from simulations using a variety of sources, as well as a simulation with no source. Simulations with sources where time averaging is performed using a short-time (long-time) window are denoted with solid (dash-dotted) lines. Simulations in the top panel are performed with the Krook operator in the boundary region, while simulations in the bottom panel are performed without this operator. }
    \label{fig:sources}
\end{figure}

In general, sources and sinks are required for global nonlinear simulations because a mismatch of flux between the two boundaries can lead to an accumulation or deficit of particles, momentum and heat in the central domain.  Our scale-separated approach to global gyrokinetics involves a novel projection operator source that is detailed in \S \ref{sec:sources}. Here, we  demonstrate the efficacy of this new source by comparing it to global simulations using a more conventional Krook-based source, as well as comparing it to  global simulations that do not employ sources.  These simulations use CBC parameters with adiabatic electrons along with $ q'' a^2 = 5$, $n'' (a^2/n)  = -1$ and $T''_\mr{i} (a^2/T_\mr{i})  = -4$. The spatial and velocity resolution for each domain is $(N_x, N_y,N_z,N_\mu, N_{v_\parallel}) = (85,85,16,8,48)$, $D_\mr{hyper} = 0.2$, and $\rho_\ast = 10^{-3}$ for the central domain. The binormal extent for all domains is $\ell_y = 20\upi \rho_\mr{i}(\rcoord_0)$ while the radial extent is $\ell_x = (6/2\upi)\ell_y$. Each boundary region is comprised of six collocation points.

The box-averaged ion temperatures fluctuations from these simulations are plotted in figure~\ref{fig:sources} as a function of time. The black solid line denotes the simulation with no sources, while the orange and blue lines denote simulations with the projection-type and Krook-type sinks, respectively. Simulations in the top panel apply the Krook operator to the three points facing the physical region with $\nu_\mr{BC} = 0.1$, while the simulations in the bottom do not. For the given simulation parameters, the turbulence is strongest in the rightmost domain, resulting in a strong right-going radial flux of heat. As a result, the simulation with no sources and sinks results in an unbounded secular decrease of temperature. On the other hand, simulations employing sources and sinks with long-windowed time averaging eventually asymptote to finite values.  
These values for the projection-type operator converge as the averaging window size is increased, and are sufficiently small to prevent profile relaxation.
In the case of the Krook-operator source, the box-averaged ion temperature eventually decays to zero, though the details on how this happens is sensitive to the precise values of $\nu_\mr{S}$ and $\tau_\mr{S}$.
 In practice, one should choose a averaging window size that is longer than any of the turbulent timescales of interests; for our global simulations, this is roughly $\Delta^{-1}(a/v_\mr{thi}) \lesssim \tau_\mr{S} \lesssim \rho_\ast^{-1}(a/v_\mr{thi})$. Thus, figure~\ref{fig:sources} indicates that both the Krook-based and projection-operator-based sources are equally effective at controlling the box-averaged temperature in the simulation domain.

\subsection{Radial heat flux profiles}

\begin{figure}
    \centering
    \includegraphics[width=\textwidth]{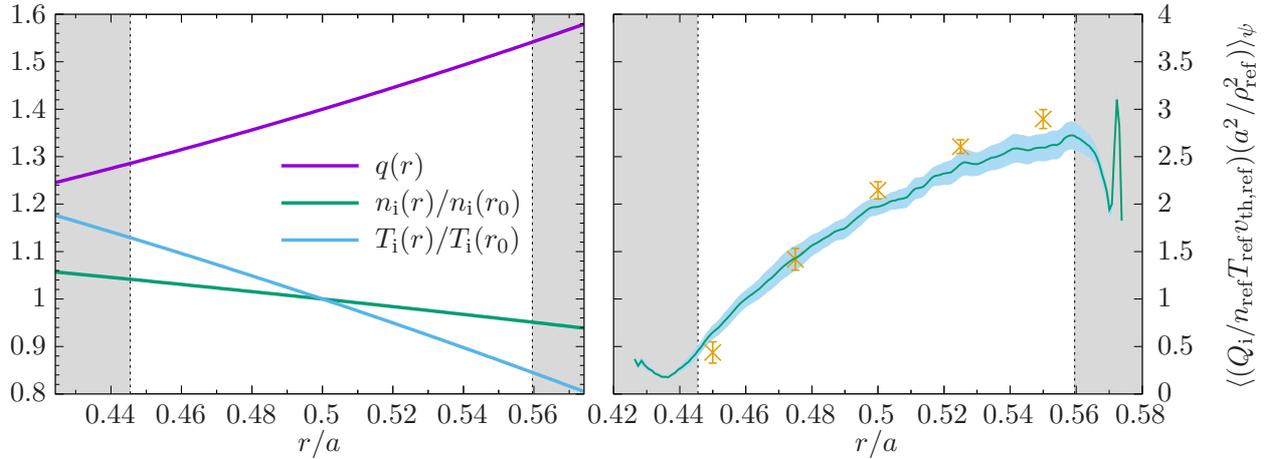}
    \caption{(\emph{left}) Radial profile of the equilibrium density and temperature, and magnetic safety factor $q$. (\emph{right}) Radial profile of the total flux-surface-averaged heat flux $Q$ (solid line) from a global $\texttt{stella}$ simulation using CBC parameters with $\rho_\ast = 10^{-3}$. Shaded areas on the left and right denote the boundary regions of the central flux tube. Yellow crosses denote local simulations performed at the specified radial location. Vertical error bars and blue shaded area encompassing the solid line are calculated using the standard error of the mean appropriate for a continuous time signal. The asymmetry of the size of the boundary regions in the physical $r/a$ space is due to $q'' \ne 0$. }
    \label{fig:flux_profile}
\end{figure}

\sloppy As a final test of our global approach to gyrokinetics, we compare the radial profile of the ion-temperature-gradient-driven nonlinear heat flux (\ref{app:fluxes})  resulting from a global simulation to the heat fluxes obtained at multiple radial positions with local simulations.  These simulations use CBC parameters with adiabatic electrons along with $ q'' a^2 = 5$, $n'' (a^2/n) = -1$ and $T''_\mr{i}(a^2/T_\mr{i}) = -4$. These parameters result in the radial profiles displayed in the left panel of figure~\ref{fig:flux_profile}. The spatial and velocity resolution for each domain is $(N_x, N_y,N_z,N_\mu, N_{v_\parallel}) = (171,85,16,8,48)$, $D_\mr{hyper} = 0.2$, and $\rho_\ast = 10^{-3}$ for the central domain. The binormal extent for all domains is $\ell_y = 20\upi \rho_\mr{i}(\rcoord_0)$ while the radial extent is $\ell_x = (6/2\upi)\ell_y$. Each boundary region is comprised of ten collocation points, with the Krook operator being applied to the six points facing the physical region with $\nu_\mr{BC} = 0.1$. The projection operator source is employed in the centre domain within the physical region with an averaging window time  of $\tau_\mr{S}(v_\mr{th,ref}/a) = 50$.

The resulting radial profile of the heat flux is plotted in the right panel of figure~\ref{fig:flux_profile}, which displays the long-time average after the turbulence has reached a saturated state. Additionally, the results of five separate local simulations are plotted using crosses.  The shaded region and error bars of the local results are calculated from the standard deviation of the mean for a continuous signal, i.e., the standard deviation divided by $\sqrt{t_\mr{ave}/t_\mr{auto}}$, where $t_\mr{ave}$ is the time window of the averaging and $t_\mr{auto}$ is the autocorrelation time of the signal. Boundary regions are denoted by the shaded areas, and the profile is given as a function of the minor radius $\rcoord$. Like the simulations from the previous section, the resulting turbulence is stronger on the right side of the simulation domain, which results in a much larger heat flux.  
Similar to the results given by~\citet{Candy_globloc}, 
the heat flux of the global simulation across the radial domain captures the radial variation of the heat fluxes resulting from by local simulations, with some minor mismatch near the boundaries. These discrepancies can be the result of a number of effects, such as profile shearing~\citep{Waltz_1994} and turbulence spreading~\cite{Garbet_1994}. With regards to the latter, the separation between any two local simulations in figure~\ref{fig:flux_profile} is approximately a single poloidal gyroradius $\rho_\mr{pol} = (B/B_\mr{P})\rho_\mr{ref}$, where $B_\mr{P}$ is the poloidal magnetic field strength, and so figure~\ref{fig:flux_profile} is consistent with turbulence spreading across eddies that have a radial extent on the order of $\rho_\mr{pol}$~\citep{Rotation_theory}. In-depth studies of both profile shearing and turbulence spreading our global approach to gyrokinetics are ongoing, and will be reported in a future publication.

\section{Conclusion}\label{sec:conclusion}

The gyrokinetic flux-tube code \texttt{stella} has been extended to allow for radially global simulations using a novel hybrid approach that retains the use of a spectral basis. While this is not the first code to take a hybrid approach to global gyrokinetics~\citep{Candy_globloc}, what makes \texttt{stella} unique are the novel radial boundary conditions that couple multiple flux-tube simulations; a subsidiary expansion of the gyrokinetic equation~\citep{Parra_globallocal} that incorporates the effects of radial profile variation in every term, rather than just $\omega_\ast$; and next-order corrections in the magnetic geometry using an extended form of the Miller equilibrium equations.  
These new features have been benchmarked using a number of standard test cases, and are readily applicable to a wide variety of physical problems, such as turbulence spreading~\citep{Garbet_1994,Lin_2004}, and the generation of intrinsic rotation due to the breaking of symmetries in the gyrokinetic equation~\citep{Rotation_theory,Wang_2010,Waltz_2011,Hornsby_2018}. We stress that while our novel radial boundary condition provides a more physically motivated alternative to the standard Dirichlet boundary condition, it does not obviate the need for Krook-type operators which smoothly connect the buffer regions to the physical one. 

Global \texttt{stella} could be usefully extended to include electromagnetic effects and arbitrary profile variation using an equilibrium solver such as EFIT~\citep{EFIT_1985}. These additions would respectively enable global simulation of high-$\beta$ tokamaks and regions of high pressure gradients such as the tokamak pedestal, the latter of which may require the resolution of  $\rho_\ast$-small physical effects with high fidelity. Finally, integrating the non-perturbative solution of quasineutrality  with \texttt{stella}'s implicit solver of the parallel streaming term  will lead to significant numerical savings when performing global simulation with kinetic electron effects, and may enable global multiscale simulations with relatively modest computational costs~\citep{Howard_2016}.

As a concluding remark, while we offer an approach to global gyrokinetics using flux-tubes, it is not the only one. Spectral methods for non-periodic domains has long been an important subject of research~\citep{Wang_1998, Agnon_1999,Boyd_2007,Huybrechs_2010}, and spectrally accurate gyroaveraging schemes for non-periodic domains have also been recently developed~\citep{guadagni_cerfon_2017}. While still in its infancy, hybrid global-local gyrokinetics will offer a physically robust and computationally efficient complement to the already vast array of global gyrokinetic codes available to the plasma physics community. 

\section*{Source code}

The \texttt{stella} source code is available at the GitHub repository~\url{https://github.com/stellaGK/stella}, while documentation can be found at ~\url{https://stellagk.github.io/stella/}.

\section*{Data availability statement}

The input files that are used for the numerical tests featured in this manuscript, as well as instructions for their use, are 
available at the following URL: \url{https://ora.ox.ac.uk/objects/uuid:68e46124-82e5-4a0b-ac78-d8f20bacdc59}. 

\section*{Acknowledgements}

The authors would like to thank the reviewers for their careful reading of the manuscript, as well as their helpful comments and suggestions. This work has been carried out within the framework of the EUROfusion Consortium and has received funding from the Euratom research and training programme 2014-2018 and 2019-2020 under grant agreement No 633053. The views and opinions expressed herein do not necessarily reflect those of the European Commission.  This work was supported in part by the Engineering and Physical Sciences Research Council (EPSRC) [Grant Number  EP/R034737/1]. The authors acknowledge EUROfusion, the EUROfusion High Performance Computer (Marconi-Fusion) under the project MULTEI and OXGK, and the use of ARCHER through the Plasma HEC Consortium [EPSRC Grant Number EP/R029148/1] under the project e607. This work was also supported by the U.S. Department of Energy under contract number DE-AC02-09CH11466. The United States Government retains a non-exclusive, paid-up, irrevocable, world-wide license to publish or reproduce the published form of this manuscript, or allow others to do so, for United States Government purposes.


\appendix

\section{Computation of geometrical coefficients using Miller equilibria \label{app:mag_geo}}

In this appendix we calculate the geometrical coefficients $B$, $\eb\bcdot \grad z$, $\eb\bcdot \grad B$, $|\grad x|^2$, $|\grad y|^2$, $|\grad x \bcdot \grad y|$,  $\eb \btimes \grad B \bcdot \grad y$, $\eb \btimes \bb{\kappa}\bcdot \grad x$ and $\eb \btimes \bb{\kappa}\bcdot \grad y$ in axisymmetric geometry using the Grad-Shafranov equation,
\begin{equation}\label{eqn:grad_app}
    R^2 \grad \bcdot \left(\frac{\grad \psi}{R^2}\right) = - 4 \upi R^2\frac{\od p}{\od \psi} - I \frac{\od I}{\od \psi}.
\end{equation}
We derive the geometrical coefficients using $(q,\alpha,\theta)$ coordinates, where $\alpha$ and $\theta$ are defined below. These coordinates can be related to the numerical coordinates $(x,y,z)$ using equations~\eqref{eqn:ycoord},~\eqref{eqn:xcoord_glob}, 
\begin{subequations}
\begin{align}
\grad x &= \frac{\psi'\rarg}{B_\mr{ref}}\grad q,\\
\grad y &= \frac{\psi'\rarg}{B_\mr{ref}}\grad \alpha,
\end{align}
\end{subequations}
along with $z = \theta$.\footnote{While the calculations in this section are done with $q$ as the radial coordinate in mind, the geometrical coefficients resulting when  $\psi$ is used as a radial coordinate can be obtained by using $\grad q = (\od q/\od \psi)\grad \psi$ where needed.} The Miller equilibrium is given by equations~\eqref{eqn:miller}.
In general, an axisymmetric magnetic field can be expressed as 
\begin{align}\label{eqn:baxi_app}
\bb{B} &=\grad \tor \btimes \grad \psi  + I(\psi) \grad \tor.
\end{align}
The safety factor $q$ is in general defined as
\begin{align}\label{eqn:app_q}
q = \frac{1}{2\pi} \int_0^{2\pi} \od \theta \, \frac{\bb{B}\bcdot \grad \tor}{\bb{B}\bcdot \grad \theta} =  \frac{I}{2\pi} \int_0^{2\pi}\od \theta\, \frac{\mathcal{J}}{R^2},
\end{align}
where we have used $\mathcal{J}^{-1} = \bb{B}\bcdot \grad \theta = \grad \theta\bcdot \grad \tor \btimes \grad \psi$ and $\bb{B}\bcdot \grad \tor = I(\psi) |\grad \tor|^2 = I(\psi)/R^2$.

Since $r$ is a flux label, $\psi(r,\theta) = \psi(r)$. To derive an expression for $\psi'$, where prime denotes differentiation with respect to $r$, from $\bb{B}\bcdot \grad \theta$ we get
\begin{align}
\bb{B}\bcdot \grad \theta = \mathcal{J}^{-1} = \psi' \grad r \bcdot \grad \theta \btimes \grad \tor,
\end{align}
or
\begin{align}
\mathcal{J} \psi' = \mathcal{J}_r,
\end{align}
where $\mathcal{J}_r^{-1} = \grad r \bcdot \grad \theta \btimes \grad \tor$. Then, using equation~\eqref{eqn:app_q} to solve  for $\psi'$,
\begin{align}\label{eqn:dpsidr_app}
 \psi' = \frac{I}{2\pi q} \int_0^{2\pi} \od \theta \, \frac{\mathcal{J}_r}{R^2}.
\end{align}
From properties of the reciprocal basis vectors~\citep{FluxCoordinates}, we also have the useful identities
\begin{subequations}
\begin{align}
|\grad r |^2 &=\frac{R^2}{ \mathcal{J}_r^2} \left[\left(\frac{\partial Z}{\partial \theta} \right)^2+ \left(\frac{\partial R}{\partial \theta }\right)^2\right], \\
|\grad \theta |^2 &=\frac{R^2}{ \mathcal{J}_r^2}\left(Z'^2 + R'^2\right), \\
\grad r \bcdot \grad \theta &=-\frac{R(r)^2}{ \mathcal{J}_r^2}\left(R' \frac{\partial R}{ \partial \theta}+ Z' \frac{\partial Z}{\partial \theta} \right),
\end{align}
\end{subequations}
and
\begin{align}
\mathcal{J}_r &= R\left(R'\frac{\partial Z}{\partial \theta} - Z' \frac{\partial R}{\partial \theta} \right).
\end{align}
 Also needed are the first partial derivatives of~\eqref{eqn:miller},
\begin{subequations} 
\begin{align}
R' &= R'_0 + \cos(\theta + \sin \theta \arcsin \delta (r)) -r \sin \theta \sin(\theta + \sin \theta \arcsin \delta (r)) (\arcsin \delta(r))',\\ 
Z' &= (\kappa' r + \kappa)\sin \theta, \\
\frac{\partial R}{\partial \theta} &= -\sin(\theta + \sin \theta \arcsin\delta (r))(1+\cos \theta \arcsin\delta(r)),\\ 
\frac{\partial Z}{\partial \theta} &= \kappa(r) r\cos \theta.
\end{align}
\end{subequations}

\subsection{Zeroth order coefficients}

We now begin calculating geometrical coefficients. From~\eqref{eqn:baxi_app},
\begin{align}
B^2 &= \frac{1}{R^2}\left( I^2(\psi) +  | \grad \psi|^2\right),
\end{align}
while the gradient of $B$ is  
\begin{align}
\grad B = \grad r \, B' + \grad \theta \frac{\partial B}{\partial \theta}.
\end{align}
The partial derivative of $B$ (and any other quantity in this appendix) with respect to $\theta$ is calculated in \texttt{stella} using second-order finite differencing. The radial derivative of $B$, on the other hand,  must be obtained by 
\begin{align}\label{eqn:bmag_diff}
B' =\frac{1}{ BR^2}\left(I I'  + \frac{1}{2} \left( |\grad \psi|^2\right)' \right) -  \frac{B R'}{R},
\end{align}
where
\begin{equation}\label{eqn:drgradpsi2dr_app}
      \left( |\grad \psi|^2\right)'  = 2\left[|\grad \psi|^2\left(\frac{R'}{R} - \frac{\mathsf{J}}{\mathcal{J}_r}\right) + \psi'^2\left(\frac{R}{\mathcal{J}_r}
    \right)^2\left(\frac{\partial R}{\partial \theta}\frac{\partial R'}{ \partial \theta} + \frac{\partial Z}{\partial \theta}\frac{\partial Z'}{\partial \theta}\right)\right],
\end{equation}
and we have defined $\mathsf{J} \doteq \psi' \mathcal{J}' = \mathcal{J}'_r - \mathcal{J}_r \psi''/\psi'$.
We can evaluate both $\mathsf{J}$ and  $I'$ using the Grad-Shafranov equation,
\begin{align}\label{eqn:grad_alt_app}
I I' &= - 4\pi R^2 p' - | \grad r |^2 \psi' \psi''  - R^2 \psi'^2 \grad \bcdot\left(\frac{\grad r}{R^2}\right),
\end{align}
where $\psi''$ can be obtained from equation~\eqref{eqn:dpsidr_app},
\begin{align}\label{eqn:ddrR2}
\psi'' =\left(\frac{I'}{I}-\frac{q'}{q}\right) \psi' +\frac{I}{2\pi q}\int_0^{2\pi}\od \theta\,  \left(\frac{\mathcal{J}_r}{R^2}\right)'.
\end{align}
The last term of~\eqref{eqn:grad_alt_app} can be readily calculated,
\begin{align}
\grad \bcdot \left(\frac{\grad r}{R^2}\right) 
&= \frac{1}{\mathcal{J}_r}\left\{ \frac{ 1}{\mathcal{J}_r} \left[\left(\frac{\partial Z}{\partial \theta} \right)^2+ \left(\frac{\partial R}{\partial \theta }\right)^2\right] \right\}' - \frac{1}{\mathcal{J}_r}\frac{\partial r_z}{\partial \theta},
\end{align}
where
\begin{subequations}
\begin{align}
r_z &\doteq\frac{1}{\mathcal{J}_r} \left(R'\frac{\partial R}{\partial \theta} + Z'\frac{\partial Z}{\partial \theta}\right),\\
\frac{\partial r_z}{\partial \theta} &= -\frac{r_z}{\mathcal{J}_r}\frac{\partial \mathcal{J}_r}{\partial \theta} + \frac{1}{\mathcal{J}_r} \left(\frac{\partial R'}{\partial \theta}\frac{\partial R}{\partial \theta} + R'\frac{\partial^2 R}{\partial \theta^2} + \frac{\partial Z'}{\partial \theta}\frac{\partial Z}{\partial \theta}+ Z'\frac{\partial^2 Z}{\partial \theta^2}\right).
\end{align}
\end{subequations}
We now eliminate $\psi''$ by combining~\eqref{eqn:grad_alt_app} and~\eqref{eqn:ddrR2},
\begin{align}
 \frac{I'}{I}  \left(1   + \frac{I^2}{|\grad \psi|^2}\right) = &- \frac{4\pi R^2}{|\grad \psi|^2}p' + \frac{q'}{q}
 - \frac{R^2}{|\grad r|^2}  \frac{1}{\mathcal{J}_r}\left[ \left(\frac{\mathcal{J}_r}{R^2} |\grad r |^2\right)'- \frac{\partial  r_z}{\partial \theta} \right] - \frac{I}{2\pi q \psi'}\int_0^{2\pi}\od \theta\, \left(\frac{\mathcal{J}_r}{R^2}\right)' \
.
\end{align}
Multiplying by $\mathcal{J}_r/R^2$, integrating over $\theta$ and using~\eqref{eqn:dpsidr_app} gives $I'$ in terms of known quantities,
\begin{align}\label{eqn:Iprime}
 \frac{I'}{I}\int_0^{2\upi} \od \theta \, \frac{\mathcal{J}_r}{R^2}\left(1   + \frac{I^2}{|\grad \psi|^2}\right) &= - \int_0^{2 \upi} \od \theta \left( \frac{4\pi \mathcal{J}_r}{|\grad \psi|^2}p' -    \frac{\mathcal{J}_r}{R^2} \left(\frac{q'}{q} + \frac{2R'}{R}\right)
  + \frac{1}{|\grad r|^2} \left\{ \frac{1}{\mathcal{J}_r} \left[\left(\frac{\partial Z}{\partial \theta} \right)^2+ \left(\frac{\partial R}{\partial \theta }\right)^2\right]'- \frac{\partial  r_z}{\partial \theta} \right\}\right),
\end{align}
where $\mathcal{J}'_r$ has cancelled. Using the above expression, we can express the remaining unknown term of the Grad-Shafranov equation \dblbrck{left hand side of~\eqref{eqn:grad_app}} in terms of known quantities;  from
\begin{align}
\grad \bcdot \left(\frac{\grad \psi}{R^2}\right)  
&= -\frac{\mathcal{J}'}{R^2\mathcal{J}_r}|\grad \psi|^2 + \frac{\psi'}{\mathcal{J}_r^2}\left[\left(\frac{\partial Z}{\partial \theta} \right)^2+ \left(\frac{\partial R}{\partial \theta }\right)^2\right]' - \frac{\psi'}{\mathcal{J}_r}\frac{\partial r_z}{\partial \theta},
\end{align}
we have
\begin{align}\label{eqn:j_mathsf}
 \mathsf{J} = \frac{R^2}{|\grad r|^2}\left\{\frac{\mathcal{J}_r}{\psi'^2}\left(4\pi p' + \frac{I I'}{R^2} \right) + \frac{1}{\mathcal{J}_r}\left[\left(\frac{\partial Z}{\partial \theta} \right)^2+ \left(\frac{\partial R}{\partial \theta }\right)^2\right]' - \frac{\partial r_z}{\partial \theta}\right\}.
\end{align}

We now express the magnetic field $\bb{B} = \grad \alpha \btimes \grad \psi$ using  Clebsch  coordinates by introducing  the variable $\alpha = \tor - q \vartheta$. Using~\eqref{eqn:baxi_app}, we find the requirement
\begin{align}
-q \frac{\partial \vartheta}{\partial \theta} \grad \theta \btimes \grad \psi = I \grad \tor.
\end{align}
Taking the scalar product with $\grad \zeta$ and integrating,
\begin{align}\label{eqn:vartheta_app}
\vartheta =\frac{I}{q} \int_0^\theta \od \theta'  \frac{\mathcal{J}}{R^2},
\end{align}
which leads to
\begin{align}
 \vartheta' =\left(\frac{I'}{I} - \frac{q'}{q}\right)\vartheta + \frac{I}{q} \int_0^\theta \od \theta'  \left(\frac{\mathcal{J}}{R^2}\right)'.
\end{align}
Thus,
\begin{align}
\grad \alpha &= \grad \tor - \left(q'\vartheta + q  \vartheta'\right)\grad r -   \frac{I}{\psi'} \frac{\mathcal{J}_r}{R^2}\grad \theta.
\end{align}
We can now readily calculate
\begin{subequations}
\begin{align}
\grad \alpha \bcdot \grad \theta &=  - \left(q'\vartheta + q \vartheta'\right)\grad r \bcdot \grad \theta-   \frac{I}{\psi'} \frac{\mathcal{J}_r}{R^2}|\grad \theta|^2, \\
\grad \alpha \bcdot \grad r &=  - \left(q'\vartheta + q \vartheta'\right)|\grad r|^2 -   \frac{I}{\psi'} \frac{\mathcal{J}_r}{R^2}\grad r \bcdot \grad \theta, \\
|\grad \alpha|^2 &=  \frac{1}{R^2}+ \left(q'\vartheta + q \vartheta'\right)^2 |\grad r|^2  + 2  \frac{I}{\psi'} \frac{\mathcal{J}_r}{R^2} \left(q'\vartheta + q \vartheta'\right)\grad r\bcdot \grad \theta + \left( \frac{I}{\psi'} \frac{\mathcal{J}_r}{R^2}\right)^2|\grad \theta|^2.
\end{align}
\end{subequations}
This allows us to calculate $k_\perp^2$,
\begin{align}
k_\perp^2 = k_\alpha^2 \left(q'^2 \vartheta_0^2|\grad r|^2 + 2q' \vartheta_0\grad \alpha\bcdot \grad r  + |\grad \alpha|^2\right),
\end{align}
where we have defined the ballooning angle $\vartheta_0 \doteq k_q / k_\alpha$ and $k_q = (\psi'(\rcoord_0)/B_\mr{ref}) k_x$. For zonal modes with $k_\alpha = 0$,
\begin{align}
k_\perp^2 &= k_q^2 q'^2 |\grad r|^2.
\end{align}
We also calculate the gradient $B$ drift, 
\begin{subequations}
\begin{align}
\grad \alpha \bcdot \eb \btimes \grad B 
  &=-  \frac{B B'}{\psi'}+ \frac{\psi'}{B}  \frac{\partial B}{\partial \theta} \left[\left(\grad \alpha \bcdot \grad r\right)\left( \grad \alpha \bcdot \grad \theta\right) - \grad r\bcdot \grad \theta |\grad \alpha|^2\right], \\
  \grad q \bcdot \eb \btimes \grad B 
 &= -\frac{q'}{\psi'} I \eb \bcdot \grad \theta \frac{\partial B}{\partial \theta},
\end{align}
\end{subequations}
and the curvature drift,
\begin{subequations}
\begin{align}
\grad q \bcdot \eb \btimes( \eb\bcdot \grad \eb) & = \grad q \bcdot \eb \btimes \frac{\grad B}{B}, \\ 
\grad \alpha \bcdot \eb \btimes( \eb\bcdot \grad \eb)  
&=  \grad \alpha \bcdot \eb \btimes \frac{\grad B}{B} + \frac{4\pi p'}{B\psi'}.
\end{align}
\end{subequations}
This completes the zeroth order in $\Delta$ calculations of the geometrical coefficients.

\subsection{First-order coefficients}

In this section we calculate the radial derivatives of the geometrical coefficients $B$, $\eb\bcdot \grad z$, $\eb\bcdot \grad B$, $|\grad x|^2$, $|\grad y|^2$, $|\grad x \bcdot \grad y|$,  $\eb \btimes \grad B \bcdot \grad y$, $\eb \btimes \bb{\kappa}\bcdot \grad x$ and $\eb \btimes \bb{\kappa}\bcdot \grad y$. These radial derivatives are needed for global simulation. Note that these coefficients are correct for arbitrary aspect ratio; previously a number of these coefficients have been calculated in the large aspect ratio limit by~\citet{Wilson_ELITE}.

To begin, note that the radial derivative of $B$ is already given by~\eqref{eqn:bmag_diff}. The radial derivative of the next two geometrical coefficients can also be readily calculated:
\begin{subequations}
\begin{align}
 \left(\eb \bcdot \grad \theta\right)' &= - \eb \bcdot \grad \theta \left(\frac{B'}{B} + \frac{\mathsf{J}}{\mathcal{J}_r}\right), \\
  \left(\eb \bcdot \grad B\right)' &=  \eb \bcdot \grad \theta \left[\frac{\partial B'}{\partial \theta} - \frac{\partial B}{\partial \theta}\left(\frac{B'}{B} + \frac{\mathsf{J}}{\mathcal{J}_r}\right) \right].
\end{align}
\end{subequations}
In order to calculate the radial derivative of the remaining terms, $\mathsf{J}'$ and $I''$ will need to be determined; this is done using the radial derivative of the equations derived from the Grad-Shafranov equation, viz.~\eqref{eqn:Iprime}, 
\begin{align}
\frac{I''}{I}\int_0^{2\upi}& \od \theta \, \frac{\mathcal{J}_r}{R^2}\left(1 + \frac{I^2}{|\grad \psi|^2}\right) + \frac{I'}{I}\int_0^{2\upi}\od \theta \,\frac{\mathcal{J}_r}{R^2}\left[\frac{I^2}{|\grad \psi|^2}\left(\frac{I'}{I} + \frac{\mathcal{J}'_r}{\mathcal{J}_r} - \frac{2R'}{R} - \frac{(|\grad \psi|^2)'}{|\grad \psi|^2}\right) +\left(\frac{\mathcal{J}'_r}{\mathcal{J}_r} - \frac{I'}{I} - \frac{2R'}{R}\right)\right] \nonumber \\
 &= \int_0^{2\upi}\od \theta \, \frac{\mathcal{J}_r}{R^2}\left[\frac{2R''}{R} - 2 \left(\frac{R'}{R}\right)^2 + \frac{q''}{q} - \left(\frac{q'}{q}\right)^2 + \left(\frac{2R'}{R} + \frac{q'}{q}\right)\left(\frac{\mathcal{J}_r'}{\mathcal{J}_r} - \frac{2R'}{R}\right)\right] \nonumber \\
 & + \int_0^{2\upi} \od \theta \left(\frac{1}{|\grad r|^2}\left\{\frac{\partial }{\partial \theta}\left[\frac{1}{\mathcal{J}_r}\left(R'\frac{\partial R}{\partial \theta} + Z'\frac{\partial Z}{\partial \theta}\right)\right] - \frac{2}{\mathcal{J}_r}\left(\frac{\partial R}{\partial \theta}\frac{\partial R'}{\partial \theta} + \frac{\partial Z}{\partial \theta}\frac{\partial Z'}{\partial \theta}\right)  \right\}\right)' \nonumber\\
 & - \int_0^{2\upi}\od \theta\, \frac{4 \upi p'\mathcal{J}_r}{|\grad \psi|^2}\left(\frac{p''}{p'} + \frac{\mathcal{J}'_r}{\mathcal{J}_r} - \frac{(|\grad \psi|^2)'}{|\grad \psi|^2}\right),
\end{align}
and~\eqref{eqn:j_mathsf},
\begin{align}
    \frac{\mathsf{J}'}{R^2} &= \frac{2 R'  \mathsf{J}}{R^3} + \frac{I\mathcal{J}_r}{R^2|\grad\psi|^2}\left[I'' + I'\left(\frac{I'}{I} + \frac{\mathcal{J}'_r}{\mathcal{J}_r} - \frac{2R'}{R} - \frac{(|\grad \psi|^2)'}{|\grad \psi|^2}\right)\right] \nonumber
    \\ &- \left(\frac{1}{|\grad r|^2}\left\{\frac{\partial }{\partial \theta}\left[\frac{1}{\mathcal{J}_r}\left(R'\frac{\partial R}{\partial \theta} + Z'\frac{\partial Z}{\partial \theta}\right)\right] - \frac{2}{\mathcal{J}_r}\left(\frac{\partial R}{\partial \theta}\frac{\partial R'}{\partial \theta} + \frac{\partial Z}{\partial \theta}\frac{\partial Z'}{\partial \theta}\right)  \right\}\right)'\nonumber
    \\ &- \frac{4 \upi p'\mathcal{J}_r}{|\grad \psi|^2}\left(\frac{p''}{p'} + \frac{\mathcal{J}'_r}{\mathcal{J}_r} - \frac{(|\grad \psi|^2)'}{|\grad \psi|^2}\right),
\end{align}
where the derivative of equation~\eqref{eqn:drgradpsi2dr_app} gives
\begin{align}
    \left(|\grad \psi|^2\right)'' &= 2 \left(\frac{R'}{R} - \frac{\mathsf{J}}{\mathcal{J}_r}\right)\left( |\grad \psi|^2\right)' + 2 |\grad \psi|^2 \left[\frac{R''}{R} - \left(\frac{R'}{R}\right)^2 - \frac{\mathsf{J}'}{\mathcal{J}_r} + \frac{\mathsf{J}\mathcal{J}'_r}{\mathcal{J}_r}\right] \nonumber \\
    &+ 2\left(\frac{R\psi'}{\mathcal{J}_r}\right)^2\left[\left(\frac{\partial R'}{\partial \theta}\right)^2 + \frac{\partial R}{\partial \theta}\frac{\partial R''}{\partial \theta} + \left(\frac{\partial Z'}{\partial \theta}\right)^2 + \frac{\partial Z}{\partial \theta}\frac{\partial Z''}{\partial \theta} + 2\left(\frac{\partial R}{\partial \theta}\frac{\partial R'}{\partial \theta}+ \frac{\partial Z}{\partial \theta}\frac{\partial Z'}{\partial \theta}\right)\left(\frac{R'}{R} - \frac{\mathsf{J}}{\mathcal{J}_r}\right)\right],
\end{align}
The quantities $\psi''$ and $\mathcal{J}'_r$, the latter given by
\begin{equation}\label{app:eqn_constraint1}
    \frac{\mathcal{J}'_r}{R} = \frac{\mathcal{J}_rR'}{R^2} + R'' \frac{\partial Z}{\partial \theta}  + R'\frac{\partial Z'}{\partial \theta}  - Z'' \frac{\partial R}{\partial \theta}- Z'\frac{\partial R'}{\partial \theta},
\end{equation}
 can be related to one-another through $\mathsf{J}$, a known quantity. For global \texttt{stella} the quantity $\psi''$ (a flux-function) is given as an input parameter provided by the user, and so~\eqref{app:eqn_constraint1} serves as a constraint in determining $R''$ and $Z''$;
 one then only needs an additional constraint to determine both $R''$ and $Z''$. We choose this constraint to take the form
 \begin{equation}\label{app:eqn_constraint2}
     R'' \frac{\partial R}{\partial \theta} +Z'' \frac{\partial Z}{\partial \theta} = 0,
 \end{equation}
 which results in
 \begin{subequations}
 \begin{align}
     R'' &= \frac{1}{|\grad r|^2}\left(\frac{R}{\mathcal{J}_r}\right)^2 \frac{\partial Z}{\partial \theta}\left[\frac{\mathcal{J}_r}{R}\left(\frac{\mathcal{J}'_r}{\mathcal{J}_r} - \frac{R'}{R}\right) - R' \frac{\partial Z'}{\partial \theta} + Z' \frac{\partial R'}{\partial \theta}\right], \\
     Z'' &= -\frac{1}{|\grad r|^2}\left(\frac{R}{\mathcal{J}_r}\right)^2 \frac{\partial R}{\partial \theta}\left[\frac{\mathcal{J}_r}{R}\left(\frac{\mathcal{J}'_r}{\mathcal{J}_r} - \frac{R'}{R}\right) - R' \frac{\partial Z'}{\partial \theta} + Z' \frac{\partial R'}{\partial \theta}\right].
 \end{align}
 \end{subequations}
 Apart from $\psi''$, both $q''$ and $p''$ must also be specified for global simulation; these are taken to be additional input parameters.
   
The next few terms radial derivatives required for global simulation are
\begin{subequations}
\begin{align}
    \left( |\grad r|^2\right)'  &= 2\left[|\grad r|^2\left(\frac{R'}{R} - \frac{\mathcal{J}'_r}{\mathcal{J}_r}\right) + \left(\frac{R}{\mathcal{J}_r}
    \right)^2\left(\frac{\partial R}{\partial \theta}\frac{\partial R'}{\partial \theta} + \frac{\partial Z}{\partial \theta}\frac{\partial Z'}{\partial \theta}\right)\right], \\
   (\grad \alpha \bcdot \grad r)' &= - \left( | \grad r|^2\right)' (\vartheta q' + q \vartheta') - |\grad r|^2 (2\vartheta' q' + q'' \vartheta + q \vartheta'') - \frac{I \mathcal{J}_r}{R^2 \psi'}\left[  (\grad r \bcdot \grad \theta)' + \grad r \bcdot \grad \theta \left(\frac{I'}{I} + \frac{\mathcal{J}'_r}{\mathcal{J}_r} - \frac{2R'}{R} - \frac{\psi'' }{\psi}\right)\right], \\
    \left(|\grad \alpha|^2\right)'  &= -\frac{2R'}{R^3} + (q' \vartheta + \vartheta' q)^2 \left( |\grad r|^2\right)'  + 2 |\grad r|^2 (q' \vartheta + \vartheta' q)(2q' \vartheta'+ \vartheta q'' + q \vartheta'') \nonumber \\
     &  +  \frac{2 I \mathcal{J}_r}{R^2 \psi'}\Bigg[ \grad r \bcdot \grad \theta (2 \vartheta' q' + \vartheta q'' + q \vartheta'') + (q'\vartheta + \vartheta' q) (\grad r \bcdot \grad \theta)'  + \grad r \bcdot \grad \theta (q' \vartheta + \vartheta' q)\left(\frac{I'}{I} + \frac{\mathcal{J}'_r}{\mathcal{J}_r} - \frac{2R'}{R} - \frac{\psi'' }{\psi}\right)\Bigg] \nonumber \\
      &  + \left(\frac{I \mathcal{J}_r}{R^2 \psi'}\right)^2\left[\left(|\grad \theta|^2\right)' + 2 |\grad \theta|^2\left(\frac{I'}{I} + \frac{\mathcal{J}'_r}{\mathcal{J}_r} - \frac{2R'}{R} - \frac{\psi'' }{\psi}\right)\right].
\end{align}
\end{subequations}
Needed for these term are 
\begin{subequations}
\begin{align}
      (\grad r \bcdot \grad \theta)'  &= 2\grad r \bcdot \grad \theta \left(\frac{R'}{R} - \frac{\mathcal{J}'_r}{\mathcal{J}}\right) - \left(\frac{R}{\mathcal{J}_r}\right)^2\left(R'' \frac{\partial R}{\partial \theta} + R' \frac{\partial R'}{\partial \theta} + Z'' \frac{\partial Z}{\partial \theta} + Z' \frac{\partial Z'}{\partial \theta}\right), \\
    \left( | \grad \theta|^2\right)' &= 2
  \left(\frac{R}{\mathcal{J}_r}\right)^2 \left[R' R'' + Z'Z'' + \left(R'^2 + Z'^2 \right)\left(\frac{R'}{R} - \frac{\mathcal{J}'_r}{\mathcal{J}_r}\right)\right],\\
    (\grad \alpha \bcdot \grad \theta)' &= -(\vartheta q' + q \vartheta')(\grad r \bcdot \grad \theta)' - \grad r \bcdot \grad \theta (2q'\vartheta' + \vartheta q'' + q \vartheta'') - \frac{I \mathcal{J}_r}{R^2 \psi'}\left[\left( |\grad \theta|^2\right)' + |\grad \theta|^2 \left(\frac{I'}{I} + \frac{\mathcal{J}'_r}{\mathcal{J}_r} - \frac{2R'}{R} - \frac{\psi''}{\psi'}\right)\right],
\end{align}
\end{subequations}
and
\begin{align}
    \vartheta'' &= \frac{I}{q}\int_0^\theta \od \theta' \frac{\mathcal{J}}{R^2}\Bigg[\left(\frac{I'}{I} + \frac{\mathsf{J}}{\mathcal{J}_r} - \frac{q'}{q} - \frac{2R'}{R}\right)^2 + \frac{I''}{I} - \left(\frac{I'}{I}\right)^2   + \frac{\mathsf{J}'}{\mathcal{J}_r} - \left(\frac{\mathsf{J}}{\mathcal{J}_r}\right)^2 - \frac{\mathsf{J}}{\mathcal{J}_r}\frac{\psi''}{\psi} - \frac{q''}{q} + \left(\frac{q'}{q}\right)^2 - \frac{2R''}{R} + 2\left(\frac{R'}{R}\right)^2  \Bigg].
\end{align}

With these, the radial derivative of the perpendicular wavenumber is given by
\begin{equation}
\left(k_\perp^2\right)' = k_\alpha^2 \left\{\left( |\grad \alpha|^2\right)' + 2 \theta_0\left[q' \left(\grad \alpha \bcdot \grad r\right)' + q'' \grad \alpha \bcdot \grad r\right] + \theta_0^2 \left[q'^2 \left(|\grad r|^2\right)' + 2 q'q'' |\grad r|^2\right]\right\}.
\end{equation}
The radial derivatives of the $\grad B$ drift terms, 
are given by
\begin{subequations}
\begin{align}
    \left(\grad q \bcdot \eb \btimes \frac{\grad B}{B}\right)' &= -\frac{I}{B \psi'}\left[q'' \eb \bcdot \grad B + q' \left(\eb \bcdot \grad B\right)' + q' \eb \bcdot \grad B \left(\frac{I'}{I} - \frac{B'}{B} - \frac{\psi''}{\psi}\right)\right], \\
    \left(\grad \alpha \bcdot \eb \btimes \frac{\grad B}{B}\right)' &= - \frac{B''}{\psi'} + \frac{\psi'' B'}{\psi'^2} + \frac{1}{B^2}\left[\frac{\partial B}{\partial \theta} (\grad \alpha \btimes \bb{B}\bcdot \grad \theta)' + \grad \alpha \btimes \bb{B}\bcdot \grad \theta\left(\frac{\partial  B'}{ \partial \theta} - \frac{2B'}{B}\frac{\partial B}{\partial \theta}\right)\right],
\end{align}
\end{subequations}
where
\begin{equation}
     (\grad\alpha \btimes \bb{B}\bcdot \grad \theta)' = \frac{\psi''}{\psi'}\grad \alpha \btimes \bb{B}\bcdot \grad \theta + \psi' \left[(\grad \alpha \bcdot \grad r)(\grad \alpha \bcdot \grad\theta)- |\grad \alpha|^2 \grad r\bcdot \grad \theta\right]'
\end{equation}
and
\begin{equation}
   B'' = \frac{1}{B R^2}\left[I'^2 + I I'' + \frac{1}{2}\left( |\grad \psi|^2\right)'' - \left(I I' + \frac{1}{2}\left( |\grad \psi|^2\right)'\right)\left(\frac{B'}{B} + \frac{2R'}{R}\right)\right] - \frac{B R'}{R}\left(\frac{R''}{R} + \frac{B'}{B} - \frac{R'}{R}\right).
\end{equation}
Likewise, the radial derivative of the curvature drift is given by
\begin{subequations}
\begin{align}
    \left(\frac{\eb}{B}\btimes \frac{\grad B}{B}\bcdot \grad \alpha\right)' &= \frac{1}{B}\left(\grad \alpha \bcdot \eb \btimes \frac{\grad B}{B}\right)' - \frac{B'}{B^2}\left(\grad \alpha \bcdot \eb \btimes \frac{\grad B}{B}\right), \\
    \left(\frac{4 \upi}{B^2}\frac{p'}{\psi'}\right)' &= \frac{4\upi p' }{B^2 \psi'}\left(\frac{p''}{p'} - \frac{\psi''}{\psi'} -  \frac{2 B'}{B}\right).
\end{align}
\end{subequations}
This completes the first order in $\Delta$ calculations of the geometrical coefficients.

\section{Moments of $g_{\bb{k},s}$ and their fluxes incorporating radial profile variation  \label{app:fluxes}}

In this section we document the expressions for various moments of the distribution function, along with their fluxes, appropriate for the expansion detailed in~\S \ref{sec:subsidiary}. The moments considered here are the fluctuating density $\delta n_s$, parallel velocity $\delta u_{\parallel s}$, and temperature $\delta T_s$, which are given in their normalized forms by
\begin{subequations}
\begin{align}
    \delta \tilde{n}_{\bb{k},s} &= \int \od \tilde{v}_\parallel \int \od \tilde{\mu}_s \, \frac{2}{\pi^{1/2}} \tilde{B} \left(  {J_{0s}} \tilde{g}_{\bb{k},s}  + \frac{Z_s}{\tilde{T}_s}\frac{F_s(\rcoord)}{F_s(\rcoord_0)}(J_{0s}^2 - 1) \rme^{-{v^2(\rcoord_0)}/v_{\mr{th}s}^2(\rcoord_0)} \tilde{\varphi}_\bb{k} \right),\\
\delta \tilde{u}_{\parallel ,\bb{k}, s} &=  \int \od \tilde{v}_\parallel \int \od \tilde{\mu}_s \, \frac{2}{\pi^{1/2}} \tilde{B}J_{0s}  \tilde{v}_\parallel   \tilde{g}_{\bb{k},s},\\
\delta \tilde{T}_{\bb{k},s} &= \frac{2}{3} \int \od \tilde{v}_\parallel \int \od \tilde{\mu}_s \tilde{B}\, \frac{2}{\pi^{1/2}} \tilde{B} \left(\tilde{v}_\parallel^2 + 2   \tilde{\mu}_s\tilde{B} - \frac{3}{2}\right)\left(  J_{0s}\tilde{g}_{\bb{k},s}  + { \frac{Z_s}{\tilde{T}_s}\frac{F_s(\rcoord)}{F_s(\rcoord_0)}(J_{0s}^2 - 1)\rme^{-{v^2(\rcoord_0)}/v_{\mr{th}s}^2(\rcoord_0)} \tilde{\varphi}_\bb{k} }\right),
\end{align}
\end{subequations}
where $J_{0s} = J_0(a_{\bb{k},s})$.
The Taylor expanding the above equations to first order yields
\begin{subequations}\label{app:moments}
\begin{align}
   \delta \tilde{n}_{\bb{k},s} &\approx \int \od \tilde{v}_\parallel \int \od \tilde{\mu}_s \, \frac{2}{\pi^{1/2}} \tilde{B} \left(  {J_{0s}} \tilde{g}_{\bb{k},s}  + \frac{Z_s}{\tilde{T}_s}(J_{0s}^2 - 1) \rme^{-{v^2(\rcoord_0)}/v_{\mr{th}s}^2(\rcoord_0)} \tilde{\varphi}_\bb{k} \right)
    \nonumber 
    \\  &\quad + \spaceOperator \Bigg\{\int \od \tilde{v}_\parallel \int \od \tilde{\mu}_s \, \frac{2}{\pi^{1/2}} \tilde{B} \Bigg[  {J_{0s}} \tilde{g}_{\bb{k},s}\left(\frac{B'}{B} + \frac{J'_{0s}}{J_{0s}}\right)  \nonumber
    \\ &\quad + \frac{Z_s}{\tilde{T}_s}(J_{0s}^2 - 1) \rme^{-{v^2(\rcoord_0)}/v_{\mr{th}s}^2(\rcoord_0)} \tilde{\varphi}_\bb{k}\left(\frac{B'}{B} + \frac{2J'_{0s}}{J^2_{0s}-1} - \frac{T'_s}{T_s} + \frac{F'_s}{F_s}\right) \Bigg]\Bigg\},\\
\delta \tilde{u}_{\parallel ,\bb{k}, s} &\approx  \int \od \tilde{v}_\parallel \int \od \tilde{\mu}_s \, \frac{2}{\pi^{1/2}} \tilde{B}J_{0s}  \tilde{v}_\parallel   \tilde{g}_{\bb{k},s} + \spaceOperator\left[ \int \od \tilde{v}_\parallel \int \od \tilde{\mu}_s \, \frac{2}{\pi^{1/2}} \tilde{B}J_{0s}  \tilde{v}_\parallel   \tilde{g}_{\bb{k},s}\left(\frac{B'}{B} + \frac{J'_{0s}}{J_{0s}}\right)\right],\\
\delta \tilde{T}_{\bb{k},s} &\approx \frac{2}{3} \int \od \tilde{v}_\parallel \int \od \tilde{\mu}_s \, \frac{2}{\pi^{1/2}} \tilde{B} \left(\tilde{v}_\parallel^2 + 2   \tilde{\mu}_s\tilde{B} - \frac{3}{2}\right)\left(  J_{0s}\tilde{g}_{\bb{k},s}  + { \frac{Z_s}{\tilde{T}_s}(J_{0s}^2 - 1)\rme^{-{v^2(\rcoord_0)}/v_{\mr{th}s}^2(\rcoord_0)} \tilde{\varphi}_\bb{k} }\right)
\nonumber
  \\  &\quad + \frac{2}{3}\spaceOperator \Bigg\{\int \od \tilde{v}_\parallel \int \od \tilde{\mu}_s \, \frac{2}{\pi^{1/2}} \tilde{B}\left(\tilde{v}_\parallel^2 + 2 \tilde{\mu}_s\tilde{B} - \frac{3}{2}\right) \Bigg[  {J_{0s}} \tilde{g}_{\bb{k},s}\left(\frac{B'}{B} + \frac{J'_{0s}}{J_{0s}} + \frac{2\tilde{\mu}_s\tilde{B}'}{\tilde{v}_\parallel^2 + 2   \tilde{\mu}_s \tilde{B} -3/2}\right)  \nonumber 
  \\ &\quad + \frac{Z_s}{\tilde{T}_s}(J_{0s}^2 - 1) \rme^{-{v^2(\rcoord_0)}/v_{\mr{th}s}^2(\rcoord_0)} \tilde{\varphi}_\bb{k}\left(\frac{B'}{B} + \frac{2J'_{0s}}{J^2_{0s}-1} - \frac{T'_s}{T_s} + \frac{F'_s}{F_s}+ \frac{2\tilde{\mu}_s\tilde{B}'}{\tilde{v}_\parallel^2 + 2   \tilde{\mu}_s \tilde{B} - 3/2}\right) \Bigg]\Bigg\},
\end{align}
\end{subequations}
where the coefficients that depend on space are evaluated at $\rcoord = \rcoord_0$ and 
\begin{equation}
    \frac{F'_s}{F_s} = \frac{n'_s}{n_s} + \frac{T'_s}{T_s}\left(\frac{E}{T_s} - \frac{3}{2}\right) - \frac{\mu_s B'}{T_s}.
\end{equation} 
Equations~\eqref{app:moments} are correct to first order in $\Delta$.

The relevant fluxes that accompany these moments are the flux of particles $\Gamma_s$, toroidal angular moment $\Pi_s$, and energy $Q_s$, which are given in their normalized flux-surface-averaged forms by
\begin{subequations}\label{app:fullfluxes}
\begin{align}
\ba{\tilde{\Gamma}_s}_\psi &= \ba{\int \od \tilde{v}_\parallel \int \od \tilde{\mu}_s \, \frac{2}{\pi^{1/2}} \tilde{B}  \left({ \tilde{\bb{v}}_\bb{E}} \bcdot \frac{\grad \psi}{\ba{|\grad \psi|}_{\psi_0}}\right) {\ba{\tilde{g}_s + \frac{Z_s  \ba{\tilde{\varphi}}_\bb{R}}{\tilde{T}_s}\frac{F_s(\rcoord)}{F_s(\rcoord_0)}\rme^{-{v^2(\rcoord_0)}/v_{\mr{th}s}^2(\rcoord_0)}}_\bb{r}}}_\psi,\\ 
\ba{\tilde{\Pi}_s}_\psi &= \ba{\tilde{m}_s \tilde{R}^2  \int \od \tilde{v}_\parallel \int \od \tilde{\mu}_s \, \frac{2}{\pi^{1/2}} {\tilde{B} (\tilde{\bb{v}}\bcdot \tilde{\grad} \tor)} \left({ \tilde{\bb{v}}_\bb{E}}\bcdot \frac{\grad \psi}{\ba{|\grad \psi|}_{\psi_0}}\right){\ba{\tilde{g}_s + \frac{Z_s  \ba{\tilde{\varphi}}_\bb{R}}{\tilde{T}_s}\frac{F_s(\rcoord)}{F_s(\rcoord_0)}\rme^{-{v^2(\rcoord_0)}/v_{\mr{th}s}^2(\rcoord_0)}}_\bb{r}}}_\psi,\\ 
\ba{\tilde{Q}_s}_\psi &=\ba{\frac{\tilde{m}_s}{2}  \int \od \tilde{v}_\parallel \int \od \tilde{\mu}_s \, \frac{2}{\pi^{1/2}} \tilde{B}  (\tilde{v}_\parallel^2 + 2 \tilde{\mu}_s \tilde{B} )  \left({ \tilde{\bb{v}}_\bb{E}} \bcdot \frac{\grad \psi}{\ba{|\grad \psi|}_{\psi_0}} \right){\ba{\tilde{g}_s + \frac{Z_s  \ba{\tilde{\varphi}}_\bb{R}}{\tilde{T}_s}\frac{F_s(\rcoord)}{F_s(\rcoord_0)}\rme^{-{v^2(\rcoord_0)}/v_{\mr{th}s}^2(\rcoord_0)}}_\bb{r}}}_\psi,
\end{align}
\end{subequations}
where 
$\ba{\cdots}_{\psi_0}$ is the flux-surface average performed on an infinitesimally thin flux surface located at $\rcoord = \rcoord_0$. Note that the contribution from the un-gyroaveraged $\varphi_\bb{k}$ to that would normally appear as a third term under the gyroaverages in equations~\eqref{app:fullfluxes} \dblbrck{cf. equations~\eqref{app:moments}} vanishes upon integration over $y$. The particle and heat fluxes are readily expanded, leading to
\begin{subequations}
\begin{align}
\ba{\tilde{\Gamma}_s}_\psi &\approx  \left< \int \od \tilde{v}_\parallel \int \od \tilde{\mu}_s \, \frac{2\tilde{B}}{\pi^{1/2}\langle |\tilde{\grad} \tilde{\rcoord}| \rangle_{\psi_0}}  \fourier^{-1}\left(\imag k_y \rho_\mr{ref} \tilde{\varphi}_\bb{k}\right)  \fourier^{-1}\Bigg\{ J_{0s}\tilde{g}_{\bb{k},s} + \frac{Z_s  }{\tilde{T}_s}J^2_{0s}\tilde{\varphi}_\bb{k}\rme^{-{v^2(\rcoord_0)}/v_{\mr{th}s}^2(\rcoord_0)}  \right. \nonumber \\ 
&\quad + \left.\spaceOperator\left[ J_{0s}\tilde{g}_{\bb{k},s}\left(\frac{B'}{B} + \frac{J'_{0s}}{J_{0s}}\right) + \frac{Z_s  }{\tilde{T}_s}J^2_{0s}\tilde{\varphi}_\bb{k}\rme^{-{v^2(\rcoord_0)}/v_{\mr{th}s}^2(\rcoord_0)}\left(\frac{B'}{B} + \frac{J'_{0s}}{J_{0s}} + \frac{F'_s}{F_s} - \frac{T'_s}{T_s}\right)\right]\Bigg\}_{\rcoord = \rcoord_0}\right>_\psi,\\  
\ba{\tilde{Q}_s}_\psi &\approx \left< \frac{\tilde{m}_s}{2} \int \od \tilde{v}_\parallel \int \od \tilde{\mu}_s \, \frac{2\tilde{B}  (\tilde{v}_\parallel^2 + 2 \tilde{\mu}_s \tilde{B} )}{\pi^{1/2}\langle |\tilde{\grad} \tilde{\rcoord}| \rangle_{\psi_0}} \fourier^{-1}\left(\imag k_y \rho_\mr{ref} \tilde{\varphi}_\bb{k}\right)  \fourier^{-1}\Bigg\{ J_{0s}\tilde{g}_{\bb{k},s} + \frac{Z_s  }{\tilde{T}_s}\frac{F_s(\rcoord)}{F_s(\rcoord_0)}J^2_{0s}\tilde{\varphi}_\bb{k}\rme^{-{v^2(\rcoord_0)}/v_{\mr{th}s}^2(\rcoord_0)} \right. \nonumber\\ 
&\quad + \left. \spaceOperator\left[ J_{0s}\tilde{g}_{\bb{k},s}\left(\frac{B'}{B} + \frac{2 \tilde{\mu}_s \tilde{B}'}{\tilde{v}_\parallel^2 + 2 \tilde{\mu}_s \tilde{B} } + \frac{J'_{0s}}{J_{0s}}\right) + \frac{Z_s  }{\tilde{T}_s}J^2_{0s}\tilde{\varphi}_\bb{k}\rme^{-{v^2(\rcoord_0)}/v_{\mr{th}s}^2(\rcoord_0)}\left(\frac{B'}{B} + \frac{J'_{0s}}{J_{0s}} + \frac{2 \tilde{\mu}_s \tilde{B}'}{\tilde{v}_\parallel^2 + 2 \tilde{\mu}_s \tilde{B} }  + \frac{F'_s}{F_s} - \frac{T'_s}{T_s}\right)\right]\Bigg\}_{\rcoord = \rcoord_0}\right>_\psi. 
\end{align}
\end{subequations}
The flux of toroidal angular momentum can be decomposed into two pieces, $\Pi_s = \Pi_{\parallel s} + \Pi_{\perp s}$, where
\begin{subequations}
\begin{align}
\ba{\tilde{\Pi}_{\parallel s }}_\psi &= \ba{\tilde{m}_s\int \od \tilde{v}_\parallel \int \od \tilde{\mu}_s \, \frac{2}{\pi^{1/2}}  \frac{\tilde{v}_\parallel {\tilde{I}}}{\langle |\tilde{\grad} \tilde{\rcoord}| \rangle_{\psi_0}} \fourier^{-1}\left(\imag k_y \rho_\mr{ref} \tilde{\varphi}_\bb{k}\right)  \fourier^{-1}\left( J_{0s}\tilde{g}_{\bb{k},s}\right)}_\psi, \\
\ba{\tilde{\Pi}_{\perp s}}_\psi &=-\left< \sqrt{\frac{\tilde{m}_s\tilde{T}_s}{Z_s^2}} \int \od \tilde{v}_\parallel \int \od \tilde{\mu}_s \, \frac{2}{\pi^{1/2}} 
\frac{\tilde{v}_\perp^2}{ \tilde{q}' \tilde{B}}\frac{\psi'(\rcoord_0)}{\psi'(\rcoord)} \frac{1}{\langle |\tilde{\grad} \tilde{\rcoord}| \rangle_{\psi_0}}{ \left( \vartheta_0 |\tilde{\grad} q|^2 \tilde{\psi}'^2 +  \tilde{\grad} \alpha \bcdot \tilde{\grad} q \, \tilde{\psi}'^2\right)} \times \right. \nonumber \\ &\qquad\qquad \left.\fourier^{-1}\left(\imag k_y \rho_\mr{ref} \tilde{\varphi}_\bb{k}\right)  \fourier^{-1}\left[ \frac{ J_1(a_{\bb{k},s})}{a_{\bb{k},s}}\left(\tilde{g}_{\bb{k},s}  + \frac{Z_s  }{\tilde{T}_s}\frac{F_s(\rcoord)}{F_s(\rcoord_0)}J_{0s}\tilde{\varphi}_\bb{k}\rme^{-{v^2(\rcoord_0)}/v_{\mr{th}s}^2(\rcoord_0)}\right)\right]\right>_\psi.
\end{align}
\end{subequations}
These have Taylor expansions
\begin{subequations}
\begin{align}
\ba{\tilde{\Pi}_{\parallel s }}_\psi &\approx  \left< \tilde{m}_s\int \od \tilde{v}_\parallel \int \od \tilde{\mu}_s \, \frac{2}{\pi^{1/2}}  \frac{\tilde{v}_\parallel {\tilde{I}}}{\langle |\tilde{\grad} \tilde{\rcoord}| \rangle_{\psi_0}} \fourier^{-1}\left(\imag k_y \rho_\mr{ref} \tilde{\varphi}_\bb{k}\right)  \fourier^{-1}\left\{ J_{0s}\tilde{g}_{\bb{k},s} + \spaceOperator \left[ J_{0s}\tilde{g}_{\bb{k},s}\left(\frac{J'_{0s}}{J_{0s}} + \frac{I'}{I}\right)\right]\right\}\right>_\psi, \\
\ba{\tilde{\Pi}_{\perp s}}_\psi &\approx - \left<\sqrt{\frac{\tilde{m}_s\tilde{T}_s}{Z_s^2}} \int \od \tilde{v}_\parallel \int \od \tilde{\mu}_s \, \frac{2}{\pi^{1/2}} 
\frac{\tilde{v}_\perp^2}{ \tilde{q}' \tilde{B}} \frac{1}{\langle |\tilde{\grad} \tilde{\rcoord}| \rangle_{\psi_0}}{ \left( \vartheta_0 |\tilde{\grad} q|^2 \tilde{\psi}'^2 +  \tilde{\grad} \alpha \bcdot \tilde{\grad} q \, \tilde{\psi}'^2\right)}  \fourier^{-1}\left(\imag k_y \rho_\mr{ref} \tilde{\varphi}_\bb{k}\right)\times \right. \nonumber \\ 
&  \fourier^{-1}\Bigg\{ \frac{ J_1(a_{\bb{k},s})}{a_{\bb{k},s}}\tilde{g}_{\bb{k},s}  + \spaceOperator\Bigg[\frac{J_1(a_{\bb{k},s})}{a_{\bb{k},s}}\tilde{g}_{\bb{k},s} \left(\frac{ \vartheta_0 \left(|\tilde{\grad} q|^2 \tilde{\psi}'^2 \right)'+  \left(\tilde{\grad} \alpha \bcdot \tilde{\grad} q \, \tilde{\psi}'^2\right)'}{ \vartheta_0 |\tilde{\grad} q|^2 \tilde{\psi}'^2 +  \tilde{\grad} \alpha \bcdot \tilde{\grad} q \, \tilde{\psi}'^2}-\frac{q''}{q'} - \frac{\psi''}{\psi'}\right) + \left(\frac{J_1(a_{\bb{k},s})}{a_{\bb{k},s}}\right)'\tilde{g}_{\bb{k},s}\Bigg] \nonumber
\\ &+ \frac{Z_s  }{\tilde{T}_s}\frac{J_1(a_{\bb{k},s})}{a_{\bb{k},s}}J_{0s}\tilde{\varphi}_\bb{k}\rme^{-{v^2(\rcoord_0)}/v_{\mr{th}s}^2(\rcoord_0)} + \spaceOperator \Bigg[  \frac{Z_s  }{\tilde{T}_s}\frac{ J_1(a_{\bb{k},s})}{a_{\bb{k},s}}J_{0s}\tilde{\varphi}_\bb{k}\rme^{-{v^2(\rcoord_0)}/v_{\mr{th}s}^2(\rcoord_0)}\left(\frac{J'_{0s}}{J_0s} + \frac{F'_s}{F_s} - \frac{T'_s}{T_s} - \frac{q''}{q'} - \frac{\psi''}{\psi'} \right) \nonumber
\\ &+\left. \frac{Z_s  }{\tilde{T}_s}\frac{ J_1(a_{\bb{k},s})}{a_{\bb{k},s}}J_{0s}\tilde{\varphi}_\bb{k}\rme^{-{v^2(\rcoord_0)}/v_{\mr{th}s}^2(\rcoord_0)}\frac{ \vartheta_0 \left(|\tilde{\grad} q|^2 \tilde{\psi}'^2 \right)'+  \left(\tilde{\grad} \alpha \bcdot \tilde{\grad} q \, \tilde{\psi}'^2\right)'}{ \vartheta_0 |\tilde{\grad} q|^2 \tilde{\psi}'^2 +  \tilde{\grad} \alpha \bcdot \tilde{\grad} q \, \tilde{\psi}'^2}  + \frac{Z_s  }{\tilde{T}_s}\left(\frac{ J_1(a_{\bb{k},s})}{a_{\bb{k},s}}\right)'J_{0s}\tilde{\varphi}_\bb{k}\rme^{-{v^2(\rcoord_0)}/v_{\mr{th}s}^2(\rcoord_0)}\Bigg]\Bigg\}\right>_\psi,
\end{align}
\end{subequations}
where
\begin{equation}
    \left(\frac{J_1(a_{\bb{k},s})}{a_{\bb{k},s}}\right)' = \left(\frac{J_0(a_{\bb{k},s})}{2} - \frac{J_1(a_{\bb{k},s})}{a_{\bb{k},s}}\right)\left(\frac{(k_\perp^2)'}{k^2_\perp}-\frac{B'}{B}\right).
\end{equation}
Volume-averaging can be done by integrating over the volume using the volume element $\od V =  (\psi'/q')\mathcal{J}\od \theta \od \alpha \od q $. Note that $\od V$ is also Taylor expanded, giving 
\begin{equation}
\od V \approx  \frac{\psi'}{q'}\mathcal{J}\od \theta \od \alpha \od q\left[1 + \left(r_\mr{clamped}-\rcoord_0\right)\left(\frac{\mathcal{J}'}{\mathcal{J}}+\frac{\psi''}{\psi'} - \frac{q''}{q'} \right)\right].
\end{equation}

\bibliographystyle{unsrtnat}
\bibliography{refs}

\end{document}